\documentclass[sigconf]{acmart}

\setcopyright{iw3c2w3}
\copyrightyear{2021}
\acmYear{2021}
\acmConference[WWW '21]{Proceedings of the Web Conference 2021}{April 19--23, 2021}{Ljubljana, Slovenia}
\acmBooktitle{Proceedings of the Web Conference 2021 (WWW '21), April 19--23, 2021, Ljubljana, Slovenia}
\acmPrice{}
\acmDOI{10.1145/3442381.3450001}
\acmISBN{978-1-4503-8312-7/21/04}

\usepackage[utf8]{inputenc}
\usepackage[ruled,vlined]{algorithm2e}
\usepackage{amsthm}

\newtheorem{theorem}{Theorem}

\newtheorem{lemma}[theorem]{Lemma}

\usepackage{amsmath}
\usepackage{tcolorbox}
\usepackage{amsthm,tcolorbox}

\usepackage{caption}

\newtheoremstyle{stmt}
 {3pt}
 {3pt}
 {\itshape}
 {}
 {\bfseries}
 {}
 {1em}
 {(\thmnumber{#2})}

\theoremstyle{stmt}
\newtheorem{stmt}{}[section]
\newtheorem{stmt*}[stmt]{}

\tcolorboxenvironment{stmt*}{
  frame empty,
  colback=black!15!white,
  grow to left by=6pt,
  grow to right by=6pt,
  left=1.6pt,
  right=1.6pt,
  arc=0pt
}

\begin{document}

\title{Random Graphs with Prescribed $K$-Core Sequences: \\
A New Null Model for Network Analysis}
\author{Katherine Van Koevering}
\email{kav64@cornell.edu}
\affiliation{
  \institution{Cornell University}
}

\author{Austin R.\ Benson}
\email{arb@cs.cornell.edu}
\affiliation{
  \institution{Cornell University}
}

\author{Jon Kleinberg}
\email{kleinberg@cornell.edu}
\affiliation{
  \institution{Cornell University}
}

\begin{abstract}
    In the analysis of large-scale network data, a fundamental operation is the comparison of observed phenomena to the predictions provided by null models: when we find an interesting structure in a family of real networks, it is important to ask whether this structure is also likely to arise in random networks with similar characteristics to the real ones. A long-standing challenge in network analysis has been the relative scarcity of reasonable null models for networks; arguably the most common such model has been the configuration model, which starts with a graph $G$ and produces a random graph with the same node degrees as $G$. This leads to a very weak form of null model, since fixing the node degrees does not preserve many of the crucial properties of the network, including the structure of its subgraphs. 

Guided by this challenge, we propose a new family of network null models that operate on the $k$-core decomposition. For a graph $G$, the $k$-core is its maximal subgraph of minimum degree $k$; and the core number of a node $v$ in $G$ is the largest $k$ such that $v$ belongs to the $k$-core of $G$. We provide the first efficient sampling algorithm to solve the following basic combinatorial problem: given a graph $G$, produce a random graph sampled nearly uniformly from among all graphs with the same sequence of core numbers as $G$. This opens the opportunity to compare observed networks $G$ with random graphs that exhibit the same core numbers, a comparison that preserves aspects of the structure of $G$ that are not captured by more local measures like the degree sequence. We illustrate the power of this core-based null model on some fundamental tasks in network analysis, including the enumeration of networks motifs. 
\end{abstract}

\ccsdesc[500]{Theory of computation~Graph algorithms analysis}

\keywords{k-core, motif, Markov chain}

\maketitle

\section{Introduction}

\newcommand{\xhdr}[1]{\paragraph{\bf {#1}}}

Random graphs have long played a central role in the area of
network analysis, and one of their crucial uses has been as
{\em null models}: a way of producing families of synthetic 
graphs that match observed network data on specific basic properties.
Armed with effective null models, we can take an observed network
phenomenon and ask whether a random graph with similar characteristics
would exhibit the same phenomenon or not.

This comparison to random-graph baselines is an essential
strategy, but of course the challenge is to define what we mean
by a random graph ``with similar characteristics.''
In these types of analyses, a widely-used null model --- arguably
the ubiquitous default --- is the {\em configuration model}:
given an observed network $G$, it generates random graphs sampled
uniformly at random from among all graphs with the same degree
sequence as $G$.
The configuration model has provided a powerful way of asserting
that observed properties of real networks are not simply
a consequence of the node degrees, in that they would be unlikely
in a random graph with the same degree sequence~\cite{newman2001random,fosdick2018configuring}.

Despite the widespread use of the configuration model, 
it is well-understood to be an extremely weak null model, particularly
for any question involving local rather than global structure.
In particular, a random graph with a given degree sequence will
typically have very little non-trivial local structure in the
neighborhood of any given node $v$, and very little non-trivial
community structure.
Thus, real networks will almost always look very different from 
the predictions of a random draw from the configuration model
on any question involving structures like local motifs or dense communities;
and these are some of the main questions for which people 
seek out random graphs as baselines.

Given these limitations of the configuration model, researchers have
sought other null models in which we sample uniformly or near-uniformly
over different families of graphs defined by characteristics of
a given real network.
Stanton and Pinar, for example, show how to sample from graphs that
match an observed network $G$ not just in its degree sequence but
in the pairs of degrees $(d_i, d_j)$ arising from the edges $(i,j)$ of $G$~\cite{stanton2012constructing}.
This increases the specificity of the null model, but it continues to
lack non-trivial local or community structure.
An interesting recent step toward null models designed to exhibit
local structure was taken by Orsini et al.~\cite{orsini2015quantifying}, who generalized 
and put into practice the \emph{$dK$-series} hierarchy of random graph models~\cite{mahadevan2006systematic},
where the lowest levels match the degree sequence or degree correlations
and higher-levels --- the 2.1-series and 2.5-series --- also match
statistics on triangles such as the average clustering coefficient or the sequence of clustering coefficients.
This approach comes with the obstacle, however, there are not any practical algorithms
for uniformly sampling from these subsequent levels that match more than just
degrees and pairs of degrees; as a result, while they constitute valuable
heuristics, they are not designed to provide guarantees on 
near-uniform sampling from the associated family of graphs.

Thus, a basic question has remained: given an observed graph $G$,
can we construct a null model by sampling from a family of graphs matching
characteristics of $G$ in such a way that the resulting random samples
exhibit non-trivially rich local structure and community structure?

\xhdr{The present work: A null model based on the $k$-core.}
In this paper, we provide a new approach to this question,
by showing how to uniformly sample from graphs that match $G$ 
in its {\em k-core} properties.
The resulting samples provide random-graph baselines with 
richer graph-theoretic structure than the configuration model,
and we show that they can lead to potentially different conclusions
when employed as null models.

To formulate our approach, we begin with some basic definitions.
Given a graph $G$ and a number $k$, the {\em $k$-core} of $G$ 
is the (unique) maximal subgraph of $G$ in which every node
has degree at least $k$; 
it can be found efficiently by iteratively deleting nodes of 
degree strictly less than $k$ in $G$.
(For sufficiently large $k$, $G$ will have no subgraph of 
minimum degree $k$, and hence the $k$-core of $G$ 
for these large $k$ will be the empty graph.)
Building on this definition, we say that the {\em core-value} $c_v$ of
a node $v$ is the largest $k$ such that $v$ belongs to the $k$-core of $G$.

A long line of work in network analysis has shown that successive
$k$-cores of $G$, for $k = 0, 1, 2, ...$, provides considerable
information about the local structure of $G$, including the
regions where it exhibits denser connectivity~\cite{dorogovtsev2006k,carmi2007model,shin2016corescope,laishram2018measuring,malliaros2020core}.
This information is equivalently captured by the sequence of 
core-values $c_1 \geq c_2 \geq \cdots \geq c_n$ of the $n$ nodes of $G$.

Given this, we ask the following question:
by analogy with the configuration model, which samples uniformly 
from all graphs matching the degree sequence of $G$, can we
sample uniformly (or near-uniformly) from all graphs matching
the sequence of core numbers of $G$? 
We could do this in theory by brute-force rejection sampling, so 
our goal is to develop reasonable algorithms for generating such samples.
This type of sampler would provide a genuinely new type of null model,
by producing random graphs that match an observed $G$ on 
richer forms of structure than the degree sequence.

\xhdr{Sampling a random graph with a given core-value sequence.}
We answer this question affirmatively, by providing a method
for near-uniform sampling from graphs with a given core-value sequence.
We provide an overview of our strategy here, and give details in
the subsequent sections.

Our basic approach is to define a Markov chain whose state space
is the set of all graphs with the given core-value sequence,
and whose transitions are a set of graph 
transformations that preserve the core-values.
The crux of the method, and the heart of our analysis, is the definition of a
sufficiently rich set of local transformations such that 
sequences of these transformations, composed together, 
are able to transform
a starting graph $G_0$ into any other graph with the same core-value sequence.
Applying random transformation to an underlying graph thus
produces a Markov chain on the set of all graphs of a given
core-value sequence.
Our results establish that 
the Markov chain is strongly connected; and by adding
appropriate probabilities on the ``identity transformation'' that leaves
the graph unchanged, we can also ensure that the chain is aperiodic
and has a uniform stationary distribution.
Thus, by generating random trajectories in this Markov chain,
we can sample nearly-uniformly from the set of all graphs with
a given core-value sequence.

As part of the analysis of this sampling procedure, we solve a problem
of combinatorial interest in its own right.
When we generate our Markov chain based on a given graph $G$, then
$G$ itself provides a starting state for traversing the chain.
But if we start instead from a given core-value sequence 
$c_1 \geq c_2 \geq \cdots \geq c_n$, then we face the following 
fundamental question: is the state space associated with
$(c_1, c_2, \ldots, c_n)$ non-empty?  
That is, do there exist any graphs with this core-value sequence?
And if so, can we construct one?
For degree sequences in simple graphs without loops or parallel edges,
the corresponding {\em realizability question} --- characterizing 
whether there exists a simple graph with a given degree sequence ---
is the subject of a famous theorem of Erd\"{o}s and Gallai~\cite{erdos1960graphs,choudum1986simple}
and the constructive Havel--Hakimi algorithm~\cite{havel1955remark,hakimi1962realizability}.
We provide a corresponding constructive characterization for the realizability of core-value sequences in simple graphs, 
and this gives us a starting point in the Markov chain when provided with a core-value sequence as input.

Through computational experiments, we demonstrate some of the
basic properties of the samples produced by this Markov chain, 
including how they differ systematically from the output of
the configuration model.
We then demonstrate our methods in the context of 
a {\em motif-counting} application; the question here is whether 
the frequencies of particular small subgraphs in a given graph $G$
are significantly higher, significantly lower, or indistinguishable
from the abundance of these subgraphs in a random-graph baseline.
We show that a comparison to random graphs matching the degree 
sequence of $G$ may potentially lead to different conclusions than
this same comparison to random graphs matching the core-value sequence of $G$;
this points to some of the value in having multiple null models
based on the different families of random graphs.

It is useful to note a few additional points about these results.
First, there is a large collection of additional families of random
graphs that have been studied extensively in network analysis,
including stochastic block models, preferential attachment graphs,
Kronecker graphs, and many others.
It would be interesting to relate our family of random
graphs with a given core-value sequence to these.
But there is also an important distinction to be drawn in how these
families are generally used in practice: they are typically used
as generative models specified by optimizing a constant number of
parameters and then generating graphs whose size $n$ may be arbitrarily large.
In contrast, our approach is more closely aligned with models ---
such as the configuration model and more recent approaches such as
the $dK$-series --- based on uniform or near-uniform sampling from
a family of graphs obtained by matching a base graph $G$ on
a number of parameters (such as degrees or core-values) that are
linear in the number of nodes.

Finally, we also note the following important open question.
While we prove that random walks in our Markov chain will converge
to the uniform stationary distribution on graphs of a fixed core-value
sequence, it is an open question whether this chain can be proven to
be {\em rapidly mixing}.
This question aligns in interesting ways with the fact that despite
recent progress, we still do not have a full understanding of the mixing
properties of Markov chains on graphs with fixed degree sequences either~\cite{fosdick2018configuring}.
The questions in this area are quite challenging, though 
computational evidence is consistent with the premise that
these chains tend to mix well in practice~\cite{milo2003uniform,stanton2012constructing}.
As in those cases, our computational experiments also suggest that
random walks are sampling our state space effectively in practice,
indicating the utility of our Markov-chain methods.
Establishing provable bounds is thus a valuable and potentially
quite challenging further question, and recent techniques
in the theory of rapidly mixing Markov chains might be valuable here.

\subsection{Additional related work}

There are a large number of random graph models that are used for network
analysis, and we refer to surveys by Sala et al.~\cite{sala2010measurement} and
Drobyshevskiy and Turdakov~\cite{drobyshevskiy2019random} for a more expansive
discussion.  The models most relevant to our paper are those that are employed
as ``null models,'' where the goal is to sample uniformly from the set of all
graphs satisfying a certain property and then evaluate how likely other properties are under the null.  
The \emph{configuration model},
which samples uniformly from the set of graphs with a prescribed degree sequence, is broadly used~\cite{bender1978asymptotic,bollobas1980probabilistic,molloy1995critical,molloy1998size,artzy2005generating,newman2001random,fosdick2018configuring}.
There are several variants of the configuration model for dealing with
simple graphs, self-loops, and multi-edges; these details and a host of applications are covered
in depth in the survey by Fosdick et al.~\cite{fosdick2018configuring}. 
Furthermore, there are a number of configuration-type models for other relational data models such as
hypergraphs~\cite{chodrow2020configuration} and simplicial complexes~\cite{young2017construction}.
The Chung--Lu model is similar to the configuration model but samples from from graphs whose expected
degree sequence is the same as the one that is given~\cite{chung2002average,chung2002connected,chung2004spectra}.

The space of graphs with a fixed degree sequence is a special case of the more general $dK$-graphs, which
specifies degree correlation statistics for subgraphs of size
$d$~\cite{mahadevan2006systematic} (the configuration model corresponds to $d = 1$). 
Pinar and Stanton~\cite{stanton2012constructing} developed a uniform sampler for
the $d = 2$ case, which generates graphs with a prescribed \emph{joint} degree distribution. 
Further generalizations of the $dK$-graphs include those with
prescribed degree correlations and clustering
statistics~\cite{gjoka2013,colomer2013deciphering,orsini2015quantifying}.  All
of these techniques rely on MCMC samplers, but those for the $d \geq 3$ cases or
these generalized $dK$-graphs do not guarantee uniform samples.  We also use
MCMC sampling, but we can guarantee that the stationary distribution is uniform
over the space of graphs with a specified $k$-core sequence.


A major application of null models is the determination of important small subgraph patterns,
often called \emph{network motifs}~\cite{shen2002network,milo2002network,milo2004superfamilies,sporns2004motifs,kovanen2011temporal}.
In these applications, small subgraphs are counted in the real network and the null model, and those appearing
much more or less in the data compared to the null are deemed interesting for study.
We include a set of experiments that revisits network motifs to see which are significant under our $k$-core null model.


\section{Generating Random Graphs with a Given Core-Value Sequence}

\def\c{{\bf c}}
\def\H{\mathcal H}
\def\Hc{\H_\c}
\def\S{\mathcal S}
\def\Sc{\S_\c}
\def\core{\Gamma}
\def\endpf{\rule{2mm}{2mm}}

\newcommand{\omt}[1]{}
\newcommand{\rf}[1]{(\ref{#1})}
\newcommand{\mv}[2]{\begin{itemize} \item {\em Move {#1}. {#2}} \end{itemize}}

For generating a random graph with a given core-value sequence
$\c = c_1 \geq c_2 \geq \cdots \geq c_n$, we will proceed as follows.
First, we define the {\em state space} $\Sc$ to be the set of
all graphs with core-value sequence equal to $\c$.
In this section, as in the rest of the paper,
all graphs are undirected and {\em simple}, with no self-loops
or parallel edges.

We will define a set of {\em moves} that apply to a graph $G \in \Sc$;
each move transforms $G$ into another graph $G' \in \Sc$
(where possibly $G' = G$).
The moves are defined such that if there is a move from $G$ to $G'$,
there is also one from $G'$ to $G$.
This allows us to define an undirected graph $\Hc$ on the state space $\Sc$, 
in which $G$ and $G'$ are connected by an edge (or potentially
by several parallel edges) if there is a move that transforms $G$ into $G'$.

Let $\Delta$ be the maximum number of legal moves out of any one 
$G \in \Hc$.
We now define a random walk with self-loops as follows:
For a graph $G$ with $D \leq \Delta$ legal moves out of it, the 
random walk remains at $G$ with probability $1 - D / (2 \Delta)$,
and with probability $D / 2 \Delta$, it chooses one of the 
$D$ legal moves out of $G$.

Our main technical result is to 
show that for any two graphs $G_1, G_2 \in \Sc$, it is 
possible to apply a sequence of moves that tranforms $G_1$ into $G_2$.
This means that the undirected graph $\Hc$ we have defined is connected,
and so the random walk we have defined converges from any starting point
to a unique stationary 
distribution that (by the definition of the transition probabilities) 
is uniform on $\Sc$.
We can therefore run the Markov chain from an arbitrary starting point,
and the graph we have after $t$ steps will become arbitrarily close
to a uniform graph with core-value sequence $\c$ as $t \rightarrow \infty$.

For the starting point, we can either use a given input graph,
or we can start directly from a core-value sequence $\c$ and 
construct a graph that realizes this sequence, if one exists.
We show first how to efficiently perform this latter operation, constructing a 
graph from a core-value sequence.

\subsection{The realization problem for core-value sequences}

Given a sequence $\c = c_1 \geq c_2 \geq \cdots \geq c_n$,
how can we efficiently determine if there is a graph that has this
as its core-value sequence, and to construct such a graph if one exists?
Erdos and Gallai solved the analogous problem for degree sequences
\cite{erdos1960graphs,choudum1986simple},
and here we give an efficient algorithm for core-value sequences.

Since core-values are define by degrees of subgraphs, it is useful
to have some initial terminology for degree sequences as well.
Recall that a graph is called {\em $d$-regular} if all of its
node degrees are equal to $d$.
We observe the following.

\begin{stmt}
If $d$ is an even number, there exist $d$-regular graphs on every
number of nodes $n \geq d+1$.  
If $d$ is an odd number, there exists a $d$-regular graph on $n \geq d+1$ 
nodes if and only if $n$ is even.
\label{stmt:regular-exists}
\end{stmt}
\begin{proof}
There are many natural constructions; here is one that is easy to describe.
We label the nodes $0, 1, \ldots, {n-1}$ and interpret addition 
modulo $n$ (thus imagining the nodes organized in clockwise order).
When $d$ is even, 
connect each node $i$ to the $d/2$ nodes on either side of it in this order:
${i - d/2}, i - (d/2) + 1, \ldots i + (d/2)$.
When $d$ is odd and $n$ is even, 
connect each node $i$ to the nodes 
${i - (d-1)/2}, i - ((d-1)/2) + 1, \ldots i + ((d-1)/2)$ as well as
the ``antipodal'' node in the clockwise order, $i + (n/2)$.

Finally, we note that in any graph, the sum of the degrees of all
nodes must be an even number (since every edge is counted twice),
and therefore when $d$ is odd, any $d$-regular graph must have
an even number of nodes.
\end{proof}

It will be useful to be able to talk about ``almost regular''
graphs when $d$ is odd and $n$ is odd, so we say that a graph $G$
is {\em $d$-uniform} if (i) $d$ is even and $G$ is $d$-regular;
or (ii) $d$ is odd, $G$ has an even number of nodes, and $G$ is $d$-regular;
or (iii) $d$ is odd, $G$ has an odd number of nodes, and $G$ consists
of a single node of degree $d+1$ with all other nodes having degree $d$.
By slightly extending the construction from the proof of
\rf{stmt:regular-exists} to handle case (iii) in this definition as well,
we have

\begin{stmt}
For all $d$ and all $n \geq d+1$, 
there exists a $d$-uniform graph on $n$ nodes.
\label{stmt:uniform-exists}
\end{stmt}

We now consider the set of $c$-cores of $G$, for $c = 0, 1, 2, \ldots$,
where again the $c$-core $\core_c$ is the unique
maximal subgraph of minimum degree $c$.
(In cases where it is clear from context, we will sometimes use
$\core_c$ to denote the set of nodes in the $c$-core, as well as the
subgraph itself.)
The following construction procedure for the $c$-cores of $G$ 
will be useful in the proofs as well.
\begin{itemize}
\item We first define $\core_0$ to be all of $G$.
\item Having constructed $\core_c$ for a given $c$, we then repeatedly
delete any node of degree at most
$c$ from $\core_c$, updating the degrees as we go, 
until no more deletions are possible.
(Note that while all nodes in $\core_c$ have degree at least $c$
at the start of this deletion process, some degrees in $\core_c$ might drop
below $c$ in the middle of the process.)
Once the deletions from $\core_c$ have stopped, 
all of the remaining nodes have degree at least $c+1$.
Let $H$ be this subgraph of $G$.
$H$ has minimum degree $c+1$; and since no node
deleted so far can belong to any subgraph of minimum degree $c+1$,
we see that $H$ is the unique maximal subgraph with this property.
Thus $H = \core_{c+1}$.
\item We proceed in this way until we encounter a $c$ for which 
$\core_c$ is empty; at that point, we define $c^* = c-1$, and declare
$\core_{c^*}$ to be the {\em top core} of $G$.
\item We will refer to the order in which the nodes were deleted from
$G$ in this process as a {\em core deletion order}; note that there
is some amount of freedom in choosing the order in which nodes are
deleted, and all such orders constitute valid core deletion orders.
\end{itemize}

We first consider the case in which all core-values in an $n$-node graph
$G$ are the same number $c$.
Note that in this case, we must have $n \geq c+1$, since each node
must have at least $c$ neighbors.
Conversely, as long as $n \geq c+1$, we observe that a $c$-uniform graph
on $n$ nodes has all core-values equal to $c$.
Thus we have a first realization result for core-values, for the
case where all values are the same.

\begin{stmt}
For a core-value sequence $\c = c_1 \geq \cdots \geq c_n$ where
all $c_i = c$, there exists a graph with this core-value sequence $\c$
if and only if $n \geq c+1$.
\label{stmt:realize-same-core-values}
\end{stmt}

Now, we consider an arbitrary core-value sequence 
$\c = c_1 \geq \cdots \geq c_n$.
As in \rf{stmt:realize-same-core-values}, the highest $c_1 + 1$ values
must be the same in order for node $1$ to have a sufficient number of
neighbors in the top core $\core_{c_1}$.
Thus, suppose $c_{c_1 + 1} = c_1$.

Now, suppose $|\core_{c_1}| = n_1$, where $n_1 \geq c_1 + 1$.
Let $H$ be an $n_1$-uniform graph on the nodes $1, 2, \ldots, n_1$.
For each node $j > n_1$, we attach it to an arbitrary set of $c_j$ nodes
in $H$, resulting in a graph $G$ on the nodes $1, 2, \ldots, n$.
We now claim

\begin{stmt}
The graph $G$ has core-value sequence 
$\c = c_1 \geq \cdots \geq c_n$.
\label{stmt:realize-core-value-construct}
\end{stmt}
\begin{proof}
By construction, the $n_1$ nodes $i$ with $1 \leq i \leq n_1$
all have $c_i = c_1$; they all belong to $H$ and hence have
core-value equal to $c_1$.
For $j > n_1$, note that it belongs to the subgraph induced
on the nodes $\{1, 2, \ldots, j\}$; since the minimum degree
in this subgraph is $c_j$, we have $j \in \core_{c_j}$.
But since the degree of $j$ is $c_j$, we also have
$j \not\in \core_{c_j + 1}$, and hence the core-value of $j$
is $c_j$, as required.
\end{proof}

From \rf{stmt:realize-core-value-construct} it follows that $G$
realizes the given core-value sequence $\c$.
Since the only assumption on $\c$ was that $c_{c_1 + 1} = c_1$,
we have the following theorem about realization of
core-value sequences.

\begin{stmt}
A sequence $\c = c_1 \geq \cdots \geq c_n$ is the
core-value sequence of a simple graph if and only if 
$c_{c_1 + 1} = c_1$; and when this condition holds, there
is an efficient algorithm to construct a graph with core-value sequence
equal to $\c$.
\label{stmt:realize-core-value-condition}
\end{stmt}

\subsection{A Markov Chain on All Graphs with a Given Core-Value Sequence}

In the previous subsection, we showed how to construct a single
member of the state space $\Sc$ consisting of all graphs 
with core-value sequence $\c = c_1 \geq \cdots \geq c_n$.
We now define a {\em move set} on this state space,
providing ways of transforming a given graph in $\Sc$ into other
graphs in $\Sc$.
For each move that transforms a graph $G$ to $G'$, there will also
be a move transforming $G'$ to $G$;
thus, the graph $\Hc$ on $\Sc$ in which $G$ and $G'$
are adjacent when there is a move transforming one directly into the other
is an undirected graph.

Let $G$ be a graph with core-value sequence $\c$.
We note that sorting the nodes in the decreasing
sequence of their indices $n, n-1, \ldots, 2, 1$
constitutes a core deletion order for $G$, and we will use this
fact at certain points in the analysis.

The first set of moves is

\mv{1}{Add and Delete. For any nodes $(i,j)$ not connected by an edge in $G$,
we can add the edge $(i,j)$ provided that no core-values are affected.
Similarly, for an edge $(i,j)$ of $G$, we can delete $(i,j)$ provided
that no core-values are affected.}

Given that we only add or delete edges when the core-values are unaffected,
the resulting graph $G'$ is also in $\Sc$ by definition.

The remaining moves alter multiple edges at once, while preserving
all core-values.
The second set of moves is

\mv{2}{Move Endpoint.  Let $h, i, j$ be nodes of $G$ such that 
$c_j < \min(c_h, c_i)$, with $(h,j)$ an edge of $G$ and
$(i,j)$ not an edge of $G$.
We delete $(h,j)$ and insert $(i,j)$.}

We claim

\begin{stmt}
If $G \in \Sc$ and we apply an instance of 
{\em Move Endpoint} involving nodes $h, i, j$, then the resulting
graph $G'$ is also in $\Sc$.
\label{stmt:move-endpoint-valid}
\end{stmt}
\begin{proof}
Consider the core deletion order $n, n-1, \ldots, 2, 1$ in $G$;
we consider nodes in this same order in $G'$ and analyze their core-values.
Note that $j > \max(h,i)$ since $c_j < \min(c_h, c_i)$.

First, all nodes $j' > j$ have the same edges into $\{1, 2, \ldots, j'-1\}$
in both $G$ and $G'$, so all of them will get the same core-value
and can be deleted in the same order.
Next, $j$ has the same number of edges into $\{1, 2, \ldots, j-1\}$
in both $G$ and $G'$, so it can still be deleted when we encounter
it in this order in $G'$, and it will get the same core-value as as well.
Finally, once $j$ is deleted, the subgraphs of $G$ and $G'$ induced
on the set of nodes $\{1, 2, \ldots, j-1\}$ are identical, and
so the ordering $j-1, j-2, \ldots, 2, 1$ forms a core deletion order
in both.

From this, it follows that the sequence of core-values is the same
in $G$ and $G'$, and hence the {\em Move Endpoint} operation
preserves the core-value sequence.
\end{proof}

The third set of moves is

\mv{3}{Core Collapse and Core Expand.  
Let $h, i, j$ be nodes of $G$ with $c_h > c_i$ and $c_i = c_j$.
If $(h,i)$ and $(h,j)$ are both edges of $G$ but $(i,j)$ is not,
the Core Collapse operation deletes $(h,i)$ and $(h,j)$ and inserts $(i,j)$,
provided that no core values are affected.
Analogously, if $(i,j)$ is an edge of $G$ but $(h,i)$ and $(h,j)$ are not,
the Core Expand operation deletes $(i,j)$ and inserts $(h,i)$ and $(h,j)$,
again provided that no core values are affected.

We will also allow ``half-move'' versions of Core Collapse and Core Expand,
again only in the case where no core values are affected:
in the half-move version of Core Collapse, only one of $(h,i)$ or $(h,j)$
is deleted; and in the half-move version of Core Expand,
only one of $(h,i)$ or $(h,j)$ is inserted.
}

This concludes the description of the moves.
We now analyze their global properties in the state space $\Sc$.

\subsection{Connectivity of the State Space}

Recall that our strategy is to use the set of moves specified
in the previous subsection to define an undirected graph $\Hc$ on the state
space $\Sc$ of all graphs with core-value sequence $\c$.
We now show that $\Hc$ is connected --- that is, 
for any graphs $G_1, G_2 \in \Sc$, there is a sequence of moves that
transforms $G_1$ into $G_2$.
If we then define a random walk on $\Hc$ with each edge out of a
given state chosen uniformly, and self-loop probabilities at each state
set as at the start of the section,
the resulting process is connected and aperiodic, with a uniform stationary
distribution that it converges to from any starting point.

It therefore remains only to establish the connectivity of $\Hc$.
To do this, we consider two arbitrary graphs $G_1$ and $G_2$ in $\Sc$,
and we describe a path connecting $G_1$ and $G_2$ in $\Hc$.
In order to do this, it is useful to recall a small amount of terminnology:
the {\em top core}, as before, consists of the nodes with the
highest core-value $c_1$.
Suppose that there are $n_1$ such nodes;
that is, $c_{n_1} = c_1$ and $c_{n_1 + 1} < c_1$.
Let $V_1 = \{1, 2, \ldots, n_1\}$ be the set of nodes in the top core.
Finally, for simplicity of exposition, we will assume for most
of this discussion that $c_1 > 2$.
This condition applies to all the intended applications of our methods, since
graphs with $c_1 \leq 2$ are much simpler in structure than
the networks we work with in general.
Moreover, the assumption $c_1 > 2$ can be removed with additional work;
at the end of the section we describe how to achieve analogous results for
the remaining cases of $c_1 = 2$ and $c_1 = 1$.

We construct the path from $G_1$ to $G_2$ in a sequence of steps.
Since all of our moves have analogues that perform them in the
``reverse'' direction, we can describe the construction of this path
working simultaneously
from both its endpoints at $G_1$ and $G_2$.

\xhdr{Step 1: Linking all edges to the top core.}
We first apply a sequence of moves to $G_1$ designed to produce
a graph $G_1'$ that has the same core-value sequence $\c$, 
in which all edges have at least one end in the set $V_1$.

For a number $c$, 
we use $\core_c$ as before to denote the $c$-core.
We consider the nodes following the order of a core deletion sequence
$n, n-1, \ldots, 2, 1$.
When we get to a node $i$, it has degree $c_i$ by the definition
of a core elimination sequence.
If $c_i < c_1$, then 
we consider each of $i$'s incident edges $(i,j)$ in turn, and process
this edge according to the following set of cases.
\begin{itemize}
\item If $c_j = c_1$, then we do not need to do anything, since
the edge $(i,j)$ already has one end in the top core $V_1$.
\item If $c_1 > c_j > c_i$, then we apply Move Endpoint
to delete $(i,j)$ and replace it with an edge $(h,i)$ 
for any node $h \in V_1$ that is not currently a neighbor of $i$.
Such a node $h$ must exist since $|V_1| \geq c_1 + 1$ while
the degree of $i$ is $c_i < c_1$.
By \rf{stmt:move-endpoint-valid}, all core-values are preserved
by this operation.
\item If $c_j = c_i$ and 
the degree of node $j$ is equal to $c_j$, then we apply the 
full version of the Core Expand operation, replacing the edge
$(i,j)$ with two edges $(h,i)$ and $(h,j)$ to any node $h \in V_1$
that is not a neighbor of either.
(By applying a sequence of Move Endpoint operations prior to this
Core Expand operation, we can ensure that there is at least one
node $h \in V_1$ that is not a neighbor of either $i$ or $j$.)
We claim that $i$ and $j$ still have core-values equal to $c_i$
after this operation: their core-values are at least $c_i$ since
the nodes in $\core_{c_i}$ still have minimum degree $c_i$;
and their core-values are at most $c_i$ since their degrees are
equal to $c_i$.
Since all other nodes have the same core-values before and after
this operation, the core-value sequence of the graph has been preserved.
\item If $c_j = c_i$ and 
the degree of node $j$ is greater than $c_j$, then we apply the
half-move version of the Core Expand operation, replacing the edge
$(i,j)$ with the single edge $(h,i)$ for any node $h \in V_1$
that is not a neighbor of $i$.
In this case too we claim that that $i$ and $j$ still have core-values 
equal to $c_i$ after this operation.
As before, their core-values are at least $c_i$ since
the nodes in $\core_{c_i}$ still have minimum degree $c_i$.
The core-value of $i$ is at most $c_i$ since its degree is $c_i$.
The core-value of $j$ is at most $c_j$ since we have removed an
edge incident to it, which cannot raise its core-value.
Since all other nodes have the same core-values before and after
the operation, the core-value sequence of the graph has been preserved
in this case as well.
\end{itemize}
We apply this process to each edge incident to node $i$ in turn;
and we proceed node-by-node
through the core deletion sequence in this way.

At the end of this procedure, we have the desired 
graph $G_1'$: it has the same core-value sequence $\c$, 
and all its edges have at least one end in the set $V_1$.
We apply the same process to $G_2$ as well, producing a graph $G_2'$
that also has the property that every edge has at least one end in $V_1.$

\xhdr{Step 2: Converting the top core to a $c_1$-uniform graph.}
Starting from $G_1'$, we next apply a sequence of moves so that
the edges with at least one end outside of $V_1$ remain the same,
but the subgraph induced on $V_1$ becomes $c_1$-uniform.
Note that this will preserve the core-value sequence, since 
all nodes in $V_1$ in will still have core-value equal to $c_1$.
It will also uniquely determine the degree sequence of $V_1$, since
the degree sequence of a $d$-uniform graph graph on $n$ nodes
is uniquely determined by $d$ and $n$: it consists entirely of
the value $d$ when at least one of $d$ or $n$ is even; and it consists
of a single instance of $d+1$ and all other values equal to $d$
when both $d$ and $n$ are odd.

To make the subgraph on $V_1$ $c_1$-uniform, it suffices to apply
a sequence of moves resulting in the following property:
\begin{quote}
{\em ($\ast$) Either (i) all degrees in the subgraph induced on
$V_1$ are equal to $c$, or (ii) one node in the subgraph on $V_1$
has degree $c+1$, and all others have degree $c$.}
\end{quote}
An extension of our point in the previous paragraph is the following:
which of cases (i) or (ii) occurs is determined by $c_1$ and $n_1$:
since the sum of the degrees of all nodes in the subgraph on $V_1$ must
be even, we will be in case (i) when at least one of $c_1$ or $n_1$
is even, and otherwise we will be in case (ii).

To achieve property $(\ast)$ starting from $G_1'$, we first delete
any edge if it joins two nodes $i$ and $j$ in $V_1$ 
that both have degree strictly greater than $c_1$.
Since $i$ and $j$ still belong to a subgraph of minimum degree $c_1$,
their core-values are still at least $c_1$; and since the deletion
of the edge can't have increased their core-values, they are still
at most $c_1$ as well.
After this, we may assume that there are no edges joining any nodes
in $V_1$ where both ends have degree strictly greater than $c_1$.

Next, consider any node $h$ in $V_1$ of degree at least $c_1 + 2$.
By the transformations in the previous paragraph,
all of its neighbors have degree equal to $c_1$.
Let $S$ be this set of neighbors.
Each node in $S$ has an edge to at most $c_1 - 1$ other nodes in $S$,
and so there is at least one pair of nodes in $S$, say $i$ and $j$,
that are not joined by an edge.
We apply the following transformation:
We first add the edge $(i,j)$, and then we delete the edges 
$(h,i)$ and $(h,j)$.
After this sequence of three Add and Delete moves, 
the degrees of $i$ and $j$ remain the same, and the degree of
$h$ has been reduced by two.
Since all three nodes $h, i, j$ --- as well as all other nodes of $V_1$ ---
still have degree at least $c_1$, all core-values in $V_1$ remain $c_1$.
The final thing we must verify is that in the middle of this sequence,
after adding the edge $(i,j)$, we did not increase any core values strictly
above $c_1$, thereby taking our constructed path out of the state space $\Sc$.
To show this, suppose that after adding $(i,j)$ (thereby increasing
their degrees to $c_1 + 1$), we delete $G - V_1$ and
all nodes of degree at most $c_1$ in $V_1$.
By the guarantee from the previous paragraph that there were no edges
connecting two nodes of degree greater than $V_1$ in $G$, 
the resulting subgraph of $G$ consists of a set of isolated nodes,
together with a triangle on $\{h, i, j\}$.
By our assumption that $c_1 > 2$ (in fact, it is sufficient here that
$c_1 > 1$), no node in this subgraph has
degree greater than $c_1$, and hence the graph after the addition of
the edge $(i,j)$ continues to have an empty $(c_i + 1)$-core.

If we repeatedly apply the operation in the previous paragraph,
we arrive at a point where the subgraph on $V_1$ only has nodes of
degrees $c_1$ and $c_1 + 1$, and there are no edges between any of
the nodes of degree $c_1 + 1$.
Finally, we perform a sequence of moves to reduce the number of nodes
of degree $c_1 + 1$ to at most one.
Thus, suppose there are two nodes $h$ and $\ell$ that each have
degree $c_1 + 1$.
There are two cases to consider:
\begin{itemize}
\item[(i)]
If there is a node $i$ that is a neighbor of one of $h, \ell$ but not the
other --- say that $i$ is a neighbor of $h$ but not $\ell$ --- then
we add the edge $(i,\ell)$ followed by deleting the edge $(h,i)$.
After doing this, $h$ has degree $c_1$ and $\ell$ has degree $c_1 + 2$;
by applying the procedure in the previous paragraph, we can then
reduce the degree of $\ell$ to $c_1$ while preserving all other
node degrees.
In this way, we have strictly reduced the number of nodes of degree 
$c_1 + 1$.
\item[(ii)]
Suppose that the neighbor sets of $h$ and $\ell$ in $V_1$ are the same.
Let $T$ be this set of common neighbors of $h$ and $\ell$.
We have $|T| = c_1 + 1$, each node in $T$ has degree $c_1$, and
for each node, two of its edges go to $h$ and $\ell$, so at most
$c_1 - 2$ edges go to other nodes in $T$.
Thus there is a pair of nodes in $T$, say $i$ and $j$,
that are not joined by an edge.
We add the edge $(i,j)$ and then delete the edges $(h,i)$ and $(j,\ell)$;
as above, this preserves all core-values after each move, and
strictly reduces the number of nodes of degree $c_1 + 1$.
\end{itemize}
Since we can apply at least one of these two cases to strictly
reduce the number of nodes of degree $c_1 + 1$ whenever
the number of such nodes is at least two,
we can iteratively perform this reduction until the number of
nodes of degree $c_1 + 1$ is at most one.

We have therefore arrived at the desired outcome: 
a graph $G_1''$ that agrees with $G_1'$ on all edges not contained
entirely in $V_1$, and with the property that the subgraph on 
$V_1$ is $c_1$-uniform.
We perform the same process on $G_2'$, arriving at a graph $G_2''$
whose subgraph on $V_1$ is also $c_1$-uniform.

\xhdr{Step 3: Transforming one $c_1$-uniform top core into another.}
For a set of nodes $S$ in a graph $G$, let $G[S]$ denote the
subgraph of $G$ induced on $S$.
Since the subgraphs $G_1''[V_1]$ and $G_2''[V_1]$ are both $c_1$-uniform,
their multisets of degrees are the same.
If each contains a node of degree $c_1 + 1$, 
we choose an arbitrary bijection $\pi$ from $\{1, 2, \ldots, n_1\}$
to itself that maps the node of degree $c_1 + 1$ in $G_1''[V_1]$
to the node of degree $c_1 + 1$ in $G_2''[V_1]$.
Henceforth we can take this bijection as implicit, and assume
for simplicity that the node of degree $c_1 + 1$ (if any) is
the same in $G_1''[V_1]$ and $G_2''[V_1]$.

Since the degree sequences of $G_1''[V_1]$ and $G_2''[V_1]$
are the same, it is known via results on the {\em switch chain}
\cite{fosdick2018configuring} that we can transform one
of these subgraphs into the other by a sequence of moves of the
following form: find four nodes $\{h, i, j, k\}$ for which
$(h,i)$ and $(j,\ell)$ are edges but $(h,j)$ and $(i,\ell)$ are not,
and replace the edges $(h,i)$ and $(j,\ell)$ with $(h,j)$ and $(i,\ell)$.
In our move set we do not have this operation available as a single move,
but we can accomplish it by first adding the edges
$(h,j)$ and $(i,\ell)$ and then deleting the edges 
$(h,i)$ and $(j,\ell)$.
As before, we simply need to verify that in the middle of this
sequence of two Add operations and two Delete operations,
we do not cause any nodes to achieve a core-value greater than $c_1$.
To establish this, suppose that
after the two Add operations, we delete all nodes outside $V_1$ 
together with all nodes in $V_1$ of degree $c_1$.
The only nodes remaining are the four nodes $\{h, i, j, k\}$
together with the node $m$ of degree $c_1 + 1$ (if there is one),
and the edges $(h,i)$, $(j,\ell)$, $(h,j)$, and $(i,\ell)$,
as well as any edges between $\{h, i, j, k\}$ and $m$.
Since this 5-node subgraph is not the complete graph $K_5$ 
(since it lacks the edges $(h,\ell)$ and $(i,j)$), it has an empty 4-core.
By our assumption that $c_1 > 2$, this means that there is no
subgraph of minimum degree $c_1 + 1$ after deleting all nodes of
degree at most $c_1$, and hence no node acquires a core-value
of greater than $c_1$ via our sequence of moves.

By applying a sequence of these switch moves, implemented
as sequences of two Add moves and two Delete moves each,
we can thus produce a graph $G_1^o$ that agrees with $G_1''$ on 
all edges with at least one end outside $V_1$, and such that the subgraphs
$G_1^o[V_1]$ and $G_2''[V_1]$ are isomorphic.

\xhdr{Step 4: Concatenating the Subpaths.}
The graphs $G_1^o$ and $G_2''$ are almost the same: their
induced subgraphs on $V_1$ are isomorphic, and for each node $j > n_1$,
the node $j$ has degree $c_j$ in both, with all $c_j$ edges going to
nodes in $V_1$.
The ends of these $c_j$ edges from $j$ to $V_1$ might be different
in $G_1^o$ and $G_2''$, but by applying a sequence of Move Endpoint
operations, we can shift the endpoints of $j$'s edges
to $V_1$ so that they become the same in the two graphs.
Applying such operations to every $j > n_1$, we can thus 
transform $G_1^o$ to $G_2''$ by a sequence of Move Endpoint operations
for the edges from each node $n_1 + 1, n_2 + 2, \ldots, n$ into $V_1$.

Finally, we can 
concatenate all the subpaths in $\Hc$ that we have defined using
our set of moves.  This concatenation provides
the path from $G_1$ to $G_2$ in $\Hc$: it goes via 
the intermediate graphs 
$$G_1, G_1', G_1'', G_1^o, G_2'', G_2', G_2$$
and the paths between each consecutive pair of graphs on this list
using the sequences of moves describes in this subsection.

Recall from the beginning of this section that if $D(G)$
is the number of moves out of a graph $G \in \Sc$, and
$\Delta = \max_{G \in \Sc} D(G)$, we define 
a uniform random walk on the graph $\Hc$ in which the self-loop
probability at $G$ is $1 - D(G) / (2 \Delta)$.
We have thus established that

\begin{stmt}
The graph $\Hc$ defined by our move set on the 
collection of all graphs of core-value sequence $\c$
is connected.
Moreover, the random walk on $\Hc$ based on the self-loop 
probabilities we have defined has the property that
it converges to a uniform stationary distribution from 
any starting point.
\label{stmt:walk-connected-convergence}
\end{stmt}

\xhdr{Handling the case $c_1 \leq 2$.}
As noted at the start of this subsection, the exposition
has assumed that the highest core-value $c_1$ satisfies the
assumption (mild in practice) that $c_1 > 2$.
We now show how with additional work we can remove this assumption
and still achieve comparable results.

First, consider the case in which the highest core-value $c_1$
satisfies $c_1 = 2$.
The only place in the analysis
where we use the assumption that $c_1 > 2$ is in Step 3 when 
we use two Add moves followed by two Delete moves to simulate the
single {\em switch move} that replaces two edges 
$(h,i)$ and $(j,\ell)$ with $(h,j)$ and $(i,\ell)$;
we need to ensure that no node increases its core-value when we do this.
To handle the case $c_1 = 2$, we can thus simply enhance the Markov chain
by including switch moves in the top core:
when (i) the set of four nodes $\{h, i, j, k\}$ is a subset of the top core,
(ii) $(h,i)$ and $(j,\ell)$ are edges and
(iii) $(h,j)$ and $(i,\ell)$ are not edges, then we allow a single move that
replaces the edges $(h,i)$ and $(j,\ell)$ with $(h,j)$ and $(i,\ell)$.
This preserves all core-values even when $c_1 \leq 2$.
With this extra set of moves including switch moves in the top core, 
we now have a graph $\H_\c'$ with more edges than $\H_\c$, and the analysis 
above shows that that $\H_\c'$ is connected when $c_1 = 2$.
A random walk on $\H_\c'$ is thus sufficient to generate random graphs
with a given core-value sequence when the highest core-value is 2.

Finally, the case $c_1 = 1$ has a particularly simple structure:
the core-value sequence, for some $k$, has $k$ nodes 
with core-value $1$ and $n-k$ nodes with core-value $0$.
Any $G$ with this core-value sequence 
has $n-k$ isolated nodes and $k$ nodes that
form a union of trees, each of size at least 2.
We can sample directly from this set of graphs, without recourse
to the Markov chain developed here, by adapting 
an algorithm for generating uniform spanning trees 
\cite{wilson1996generating}: we first sample from the 
size distribution of components and then sample spanning trees
of complete graphs of the chosen sizes.

\omt{
\section{Setting Up the Markov Chain}
The first step of creating a Markov chain is proving what k-core sequences are possible. This is analogous to the configuration model, where a realizability proof is required to provide a random graph with a given degree sequence. First, we prove that given a graph where all nodes have core-value $c$, we can add a new node to the graph such that this new node also has core-value $c$ without altering the core-values of the other nodes.

\begin{lemma}\label{addKNode}
Given a graph $G=(V,E)$ with $n$ nodes, such that all nodes have core-value $k*$, then it is possible to add a new node $v'$ to the graph, only modifying edges with $v'$ as an endpoint, such that all nodes in $G$ and $v'$ have core-value $k*$.
\end{lemma}

\begin{proof}
Add a node $v'$ to $G$. Attach $v'$ to an arbitrary $k*$ nodes. Since $v'$ has degree $k*$, its core-value must be $\leq k*$. Thus $v'$ cannot be part of a k-core with core-value $> k*$, and its addition will not promote any node to a k-core of core-value $> k*$. Additionally, it will not demote any node to a lower k-core. Finally, since it has $k*$ connections to nodes of core-value $k*$, it must also have core-value at least $k*$. Thus all nodes in this new graph have core-value $k*$.
\end{proof}

Thus by the argument above, it is possible to have a graph of any core-value with arbitrarily large numbers of nodes. The next step is then showing that we can add nodes with other core-values. If we assume a graph consisting of core-values $c$, then we we add any nodes such that they consist of any sequence of core-values as long as those core values are all $\leq c$.

\begin{lemma}\label{addAnyNode}
Let $G=(V,E)$ be a graph with $n$ nodes that all have core-value $k*$. Let $N$ be a set of nodes. Let $N'$ be a set of intended core-values such that each core-value $n'$ in $N'$ corresponds to a node $n$ in $N$, and all intended core values are $\leq k*$ . Then $\exists$ a graph $G'=(V', E')$ that includes all nodes from G at their current core-values, and all nodes from $N$ with their intended core values from $N'$.
\end{lemma} 
\begin{proof}
Let $N$ be the nodes that you wish to add to the graph. Sort this list of nodes by the intended core-value of the nodes, with the largest core-values first. We then add the nodes one by one. For each node you add, it either has a smaller core-value than all of the nodes in the graph, or it has a core-value equal to the minimum core-value of all the nodes in the graph. Following a similar argument to \ref{addKNode}, adding such a node will not impact the core-values of the graph. Thus you are able to add all of the nodes in $N$ one by one to get $G'$.
\end{proof}

Now that we've shown we can add nodes to a framework of same-core nodes, we need to show when such a framework can be created. In fact, we can create a graph where all nodes are of core-value $c$ if and only if there are at least $c+1$ nodes.

\begin{algorithm}\label{GraphofK}
\SetAlgoLined
\KwResult{Create a graph $G=(V,E)$ with $c$ nodes all with core-value $k$}
 Let $k > 0$ and $c>k$.\;
\uIf{$k$ == 1}{
    Connect each node to exactly one other node in a chain\;
}\uElseIf{$k$ == 2}{
    Create a circle of nodes\;
    Connect each node to it's left neighbor\;
}\uElseIf{$k$ is even and $c$ is odd, or $k$ is odd and $c$ is even}{
    Create a circle of nodes\;
    Connect each node to it's left neighbor\;
    Let $z=(c-k+1)/2$\;
    \ForEach{Node $n \in V$}{
        Attach to $n$ all nodes that are more than $z$ nodes to the left or right of $n$\;
    }
}\uElse{
    Follow the above protocol for $c=c-1$\;
    Add a node to the graph, and connect it to $k$ arbitrary nodes\;
}
 \caption{Create graph of only core-value $k$}
\end{algorithm}

We can now put this all together and show that we can create a graph from any core sequence provided that there are at least $c+1$ nodes of core value $c$ if $c$ is the largest core value.

\begin{algorithm}\label{GraphofAny}
\SetAlgoLined
\KwResult{Given a list $N$ of nodes with pre-defined core values, s.t. $|N| = c$, create a graph $G=(V,E)$ using those nodes that has no self loops.}
 Sort $N$ by the core-values, largest core-values first\;
 Let $k > 0$ be the maximum value of $N$ and $c>k$\;
 Create a graph $G'$ using Alg:\ref{GraphofK} of all the nodes of $N$ that have core-value $k$\;
 Using the strategy defined in Lemma \ref{addAnyNode}, we can then add nodes from $N$ in order\;
 \caption{Create graph of arbitrary core-values}
\end{algorithm}

Furthermore, it is clear that there cannot be any graph where the largest core value is $c$ and there are $\leq c$ nodes with that core value, so the above proofs cover all possible core value sequences.

\section{Proving connectivity of the Markov chain}
Now that we've established k-core realizability, we need a set of moves. These moves must satisfy two requirements: first, it must not change the core value sequence, and second every possible graph of this core value sequence must be reachable by some sequence of these moves. First, lets define these moves.

\subsection{Move Set}
We include several types of moves in our move set. For this section, we will refer to the core-value of node $n$ as $c_n$, the set of all nodes in core $k$ as $C_k$, and the maximum core-value as $K$. This is for labeled, undirected graphs with no self-loops.

\subsubsection{Move Endpoint}
Find an edge $(A, B)$ s.t. $c_A > c_B$. Find another node $D$ s.t. $c_B < c_D$. Delete $(A, B)$ and insert $(D, B)$. $B$'s core value will not change, since it still has the same number of connections to nodes of a greater core value. Similarly, $A$ and $D$ are unaffected by the edges.

\subsubsection{Core Collapse}
Find two edges $(A, B),  (A,D)$ s.t.$c_A > c_B$ and $c_B = c_D$, replace both of those edges with $(B, D)$. This will not affect A, as it has a larger core-value. Similarly, $B, D$ will maintain the same number of edges to nodes of core value greater than or equal to their own, so it will not affect their core values.

\subsubsection{Core Expand}
Find and edge $(A, B)$ s.t. $c_A = c_B$. Find a node $D$ s.t. $c_D > c_A$. Remove $(A,B)$. If $A$ has fewer than $c_A$ edges to nodes of greater core value, insert $(A, D)$. Similarly, if $B$ has fewer than $c_A$ edges to nodes of greater core value, insert $(B, D)$. This does not affect the core-value of $D$. It also does not affect the core-value of $A, B$ because we ensure they have only up to $c_A$ edges to nodes of higher core-values, and their number of edges do not decrease.

\subsubsection{Add}
There are a few heuristics one can use, but essentially an edge $(A,B)$ can be added only if adding that edge does not affect the core values of any nodes in the graph. This can be ensured by adding the edge, checking the core values of the new graph, and then reverting if necessary.

\subsubsection{Delete}
Slightly easier than add, you can delete an $(A,B)$ edge if both $A$ and $B$ have $c_A$ and $c_B$ edges to nodes of equal or greater core-value respectively other than $(A,B)$.

\subsubsection{Degree Collapse}
Very similar to a core-collapse, but instead done with three nodes all with core-value $K$, where one node has greater degree than the other two rather than greater core value.

\subsubsection{Degree Expand}
Very similar to a core-expand, but instead done with three nodes all core-value $K$, but such that none lose core-value, which can be checked by attempting the degree expand and then reverting if necessary.

None of these moves will alter the core value of any node in the graph, so we can guarantee that using any sequence of them will not alter the core value sequence of the graph.

\subsection{Proving Connectivity}
Now we just need to prove that any construction of a core value sequence can be reached with some sequence of these moves. To do this, we will use a graph construction we term the 'Natural State'. A graph in its 'Natural State' consists of every node of core-value $c$ having exactly $c$ edges attached to nodes of higher core-value, and no edges to any node of its same core-value, where $c < K$.

\begin{lemma}
If we omit the degree collapse and degree expand, then, ignoring nodes of core-value $K$, we can reach a particular state we call the 'Natural State'.
\end{lemma}

\begin{proof}
Suppose we have a node $n$ with core-value $c < K$. Then $n$ must have at least $c$ edges to nodes of core-value $\geq c$. If an edge goes to a higher core, we don't care about it. If it goes to a lower core, we don't care about it. If it goes to a node of core-value $c$, though, then we need to adjust the edge. Let there be such an edge, with endpoints $n, n'$ both in core $c$. We then need to apply a core expand - this will ensure that the edge is replaced by edges to higher nodes, but also that the core-values of $n, n'$ do not change. Using core expand, we can then reach the 'Natural State' of the graph.
\end{proof}

Given that we have shown it is possible to reach the 'Natural State' for any graph, all that's left is to show that we can reach some analogue of the 'Natural State' for the nodes with the highest core value. That is, we need to show that for any number of nodes $n$ with core-value $K$, using only the moves listed it is possible to reach all constructions of $n$ nodes of core-value $K$.

\begin{lemma}
The degree collapse and degree expand are enough to move any group of nodes in core $c$ to either a $c$-regular graph or a $c$-regular graph with one extra node of degree $c+1$.
\end{lemma}
\begin{proof}
Suppose you have a graph consisting only of nodes in core $c$. Each node, then, has at least $c$ edges to other nodes in the graph, and it only need $c$ edges to maintain its core value. First, find any edges between two nodes each with $deg > c$, delete those edges (this will not affect core values). Next, find a node $n$ s.t. it has degree $ \geq c+2$. All of it's neighbors must have degree $c$, which means at least two of its neighbors are not connected. Perform degree collapse on these nodes, which decreases $n$'s degree, but does not change core values or the degree of the other two nodes. Repeat for all nodes until no node has degree $ > c+1$. Repeat this again, but for nodes of degree $c+1$, which has the same guarantees. By the end of this, you will be left with either a $c$-regular graph, or a $c$-regular graph with one $c+1$ node. 
\end{proof}

Now that we have shown that there is some 'Natural State' for every graph, and an analogue state for its highest core, we can prove that using the moves above it is possible to pass from any graph $G$ through its 'Natural State' to any other construction of its core value sequence.

\begin{theorem}
The move set above is sufficient to reach every possible construction of any given core set.
\end{theorem}

\begin{proof}
We have already shown that we can reach a 'Natural State' for any graph. Using the move-endpoints, and the degree functions we can also move between all 'Natural States'. We have also shown that the top-core can be reduced to a k-regular or almost k-regular graph. Any instance of this 'regular' graph can be transformed into any other instance of a 'regular' graph using degree-stable randomization, such as described in Fosdick et al. Thus, any graph of core-values $C$ can be transformed into any other graph of core-values $C$ by travelling through the 'Natural State' and 'regular' state.
\end{proof}
}

\section{Basic set-up for doing the computational experiments}

In the previous section, we established that the Markov chain defined by the random walk on $\Hc$ will converge to a uniform stationary distribution from any starting point.  We now discuss some of the computational considerations involved in running the Markov chain so as to be able to sample from it.

\omt{
First, we need to ensure that our setup of a Markov Chain will in fact produce random samples. This requires two things: that the graph of graphs for the Markov Chain is fully connected and that each node in this graph of graphs has equal degree.

\begin{theorem}
It is possible to uniformly sample a random graph of core values $C$ by running a Markov Chain using this move-set.
\end{theorem}

\begin{proof}
We have already shown that the graph of graphs is connected. In order to uniformly sample, in that case, one simply needs to add self-loops such that the sum of actual transitions and self loops is the same for every graph in the chain. We simply overestimate the maximum number of transitions, and apply this move set many times to uniformly sample.
\end{proof}
}

The basic set up for computationally running this Markov chain has two steps. A graph is input in the form of a SparseMatrix. The core numbers are then calculated and an array of core values from largest to smallest is created. The nodes are then renamed from 1 to $n$ such that each node is distinct and the node name refers to the index of their core-value in the core array. This results in nodes named such that nodes with larger core-values have smaller names.

We then do a number of transition steps. Each transition step is identical, except for the graph being processed. The transition step takes in several values: the graph, the core array and an estimated upper bound on the highest degree of any node in the Markov chain. We then estimate an upper bound on how many possible transitions there are from this graph to other graphs. We do this by soliciting an upper bound on the number of possible transitions for each type of move - note that no two moves will ever give the same exact resultant graph. This is done by proposing many moves, not all of which are necessarily possible. We sum these upper bounds to get an upper bound on the total number of transitions from this graph. If that upper bound is larger than our estimated upper bound on the largest degree in the Markov chain, then we double that estimate and start over.

Next, we randomly select a number between 0 and the estimated degree upper bound. If the number is larger than our possible transition estimate, then we "self loop" and draw again. Otherwise, we choose proportionally randomly among the moves, and then select a random proposed move. If it is not a possible move we "self loop" and draw again. Otherwise, we apply the move, then call the transition function again. This rejection sampling method is used because calculating all possible moves is both memory intensive and time consuming.

\section{Using the Core-Value Model for Network Analysis}\label{sec:properties}

Having now established the basic method for generating random graphs with a given core-value sequence, we provide a set of computational experiments showing how it can serve as a null model for network analysis tasks, parallel to the ways in which the configuration model that fixes node degrees is used.
We will see that in some cases, the conclusions from our core-value null model form fundamental contrasts with the conclusions that would be reached using the configuration model.\footnote{Code and data for all the results in this section may be obtained from the following link: {\scriptsize \url{https://www.cs.cornell.edu/~kvank/selected_publications.html}}}

\subsection{Subgraphs and Motifs}

We begin with an application where it is natural to expect that the contrast between the configuration model and the core-value model might be apparent: in the frequency of small subgraphs.
When we are assessing the abundance of a particular subgraph in real network data, we may want to compare it to the frequency of this same subgraph in a randomized version of the network that preserves some invariant.
The configuration model, by fixing only the node degrees, destroys most of the local structure, and hence can make particular small subgraphs seem highly frequent in the real network data as a result.
Intuitively, our core-value model can be viewed as preserving enough local structure to maintain the core decomposition; will this give a different view of the abundance of small subgraphs?  
We show here that it does in general.

We begin by considering perhaps the simplest family of small subgraphs: triangles on three nodes.  After this, we move on to an analysis of small motifs more generally.  In both cases, the core-value model leads to different conclusions than the configuration model in several important respects.

\subsection*{\small Triangle-based statistics}

\begin{figure}
\includegraphics[width=0.495\columnwidth]{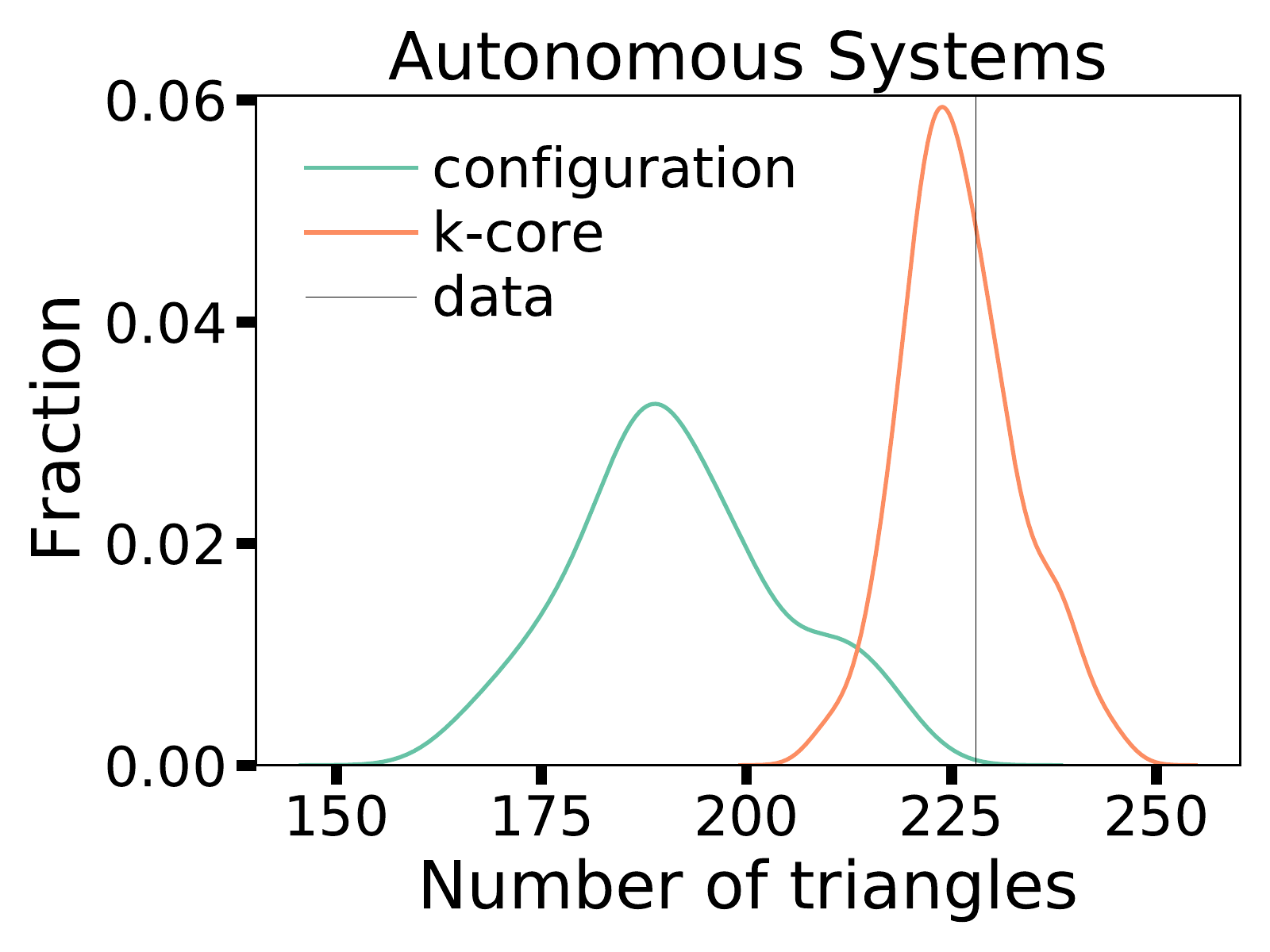}
\includegraphics[width=0.495\columnwidth]{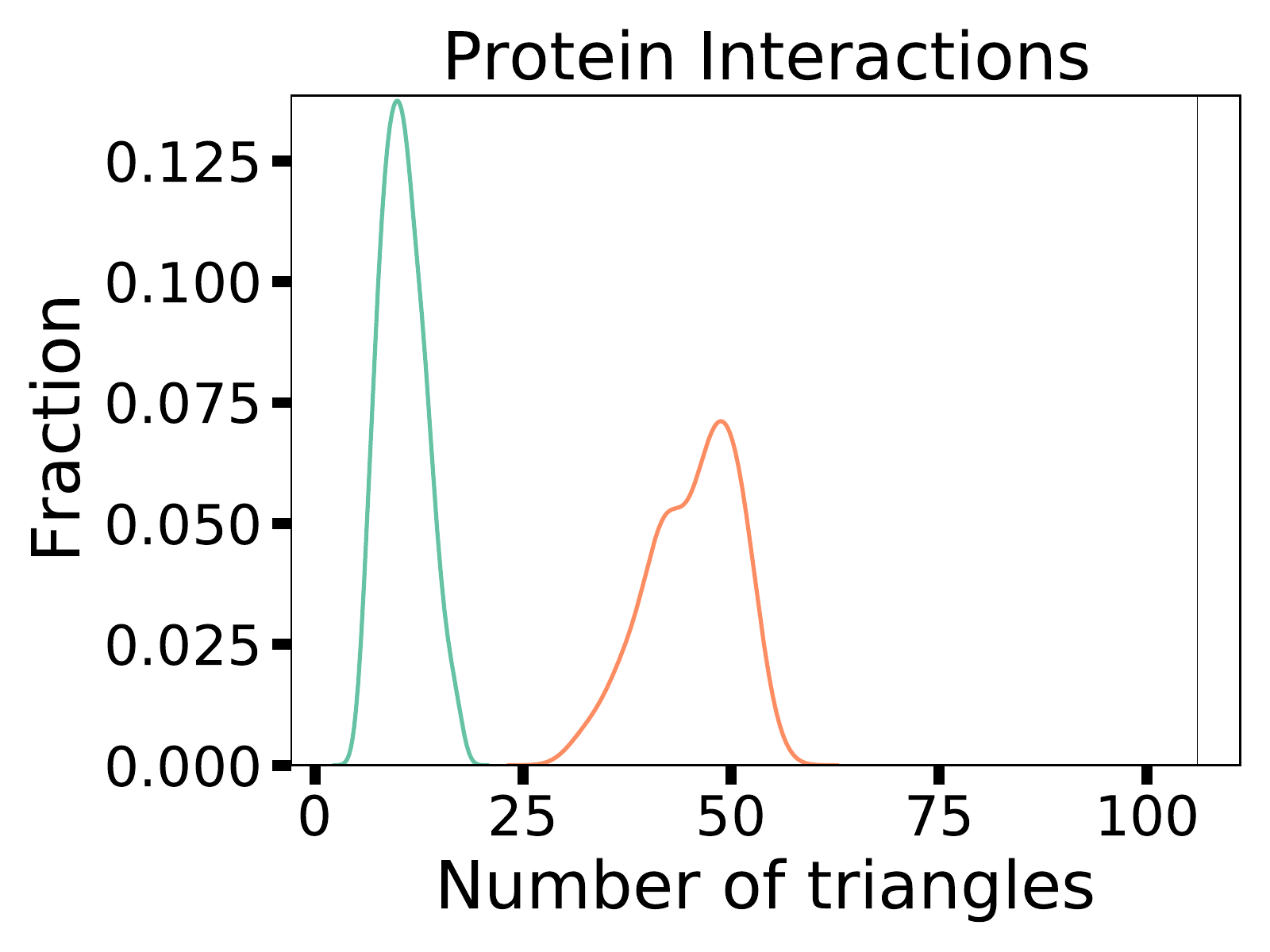}
\includegraphics[width=0.495\columnwidth]{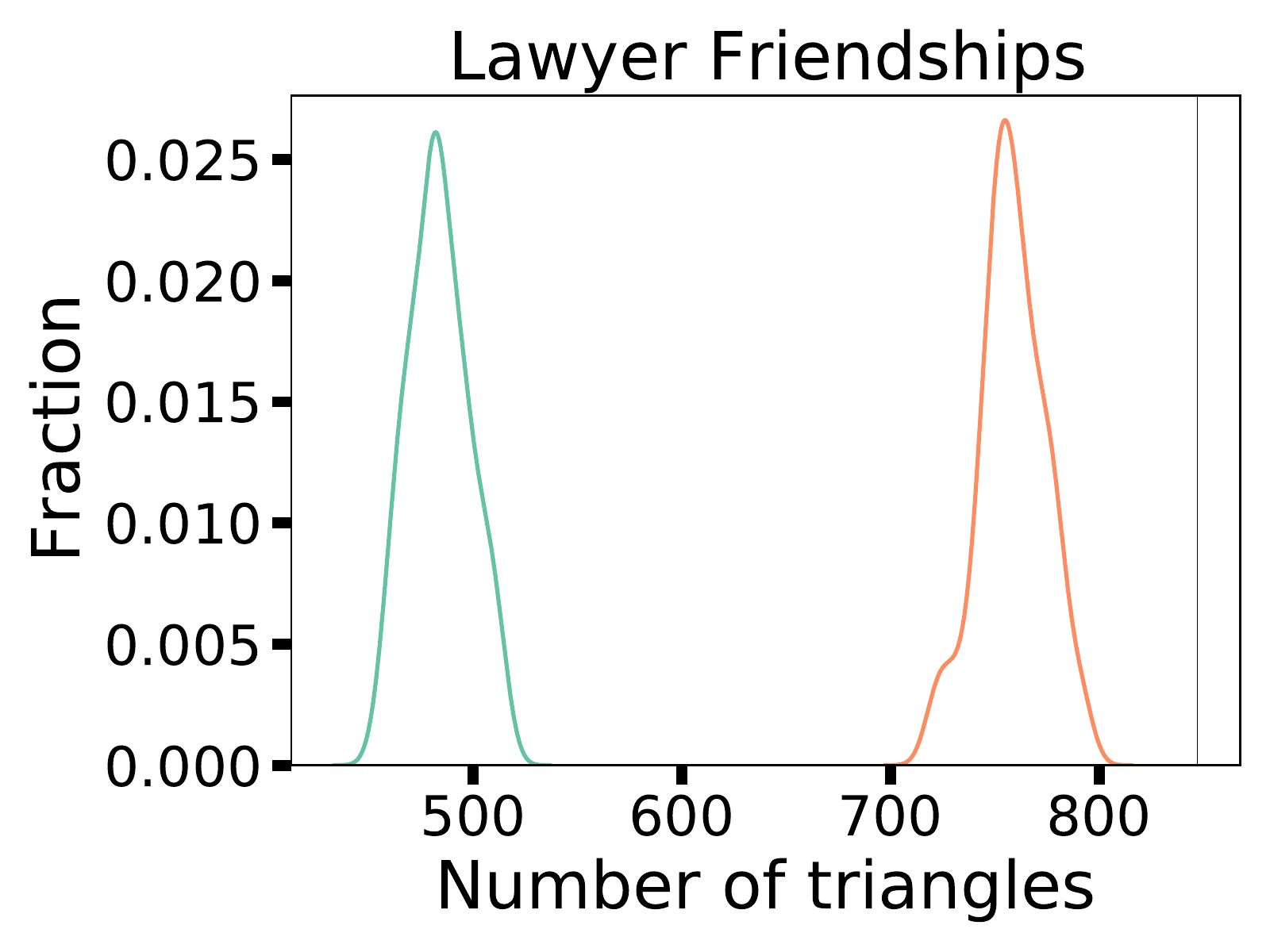}
\includegraphics[width=0.495\columnwidth]{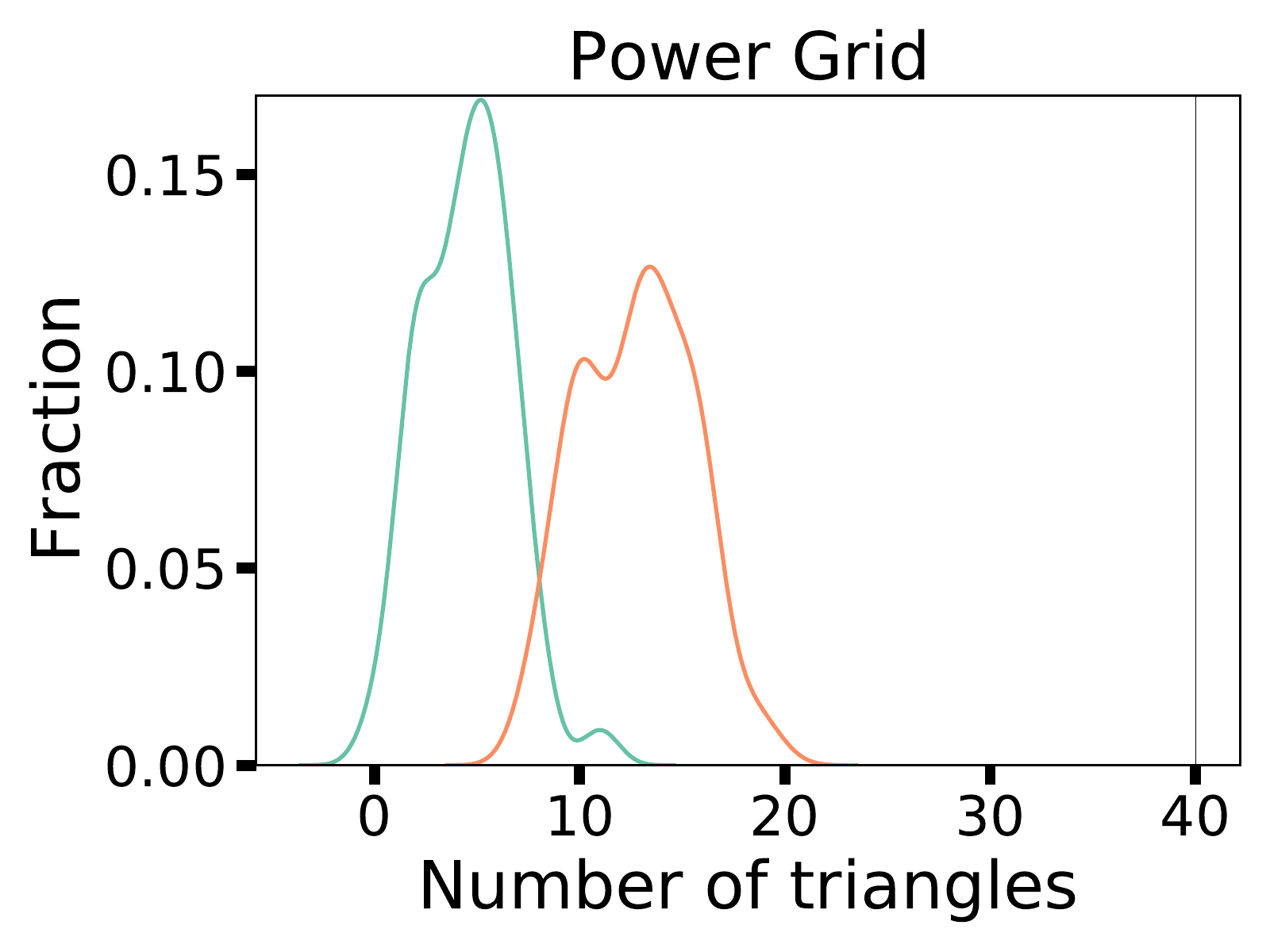}
\Description[Four histograms of triangle count]{Four triangle count histograms, the k-core null model overlaps with the configuration model for Autonomous systems and Power Grid.} 
\caption{Distribution of the number of triangles from 50 random samples of graphs with a $k$-core sequence given by a real-world graph dataset
and 50 random samples of graphs with a degree sequence given by a real-world graph dataset.
The $k$-core samples have more triangles, and often the number of triangles in the dataset is within the range those observed in the random samples.}
\label{fig:tri_dist}
\end{figure}

\begin{figure}
\includegraphics[width=0.495\columnwidth]{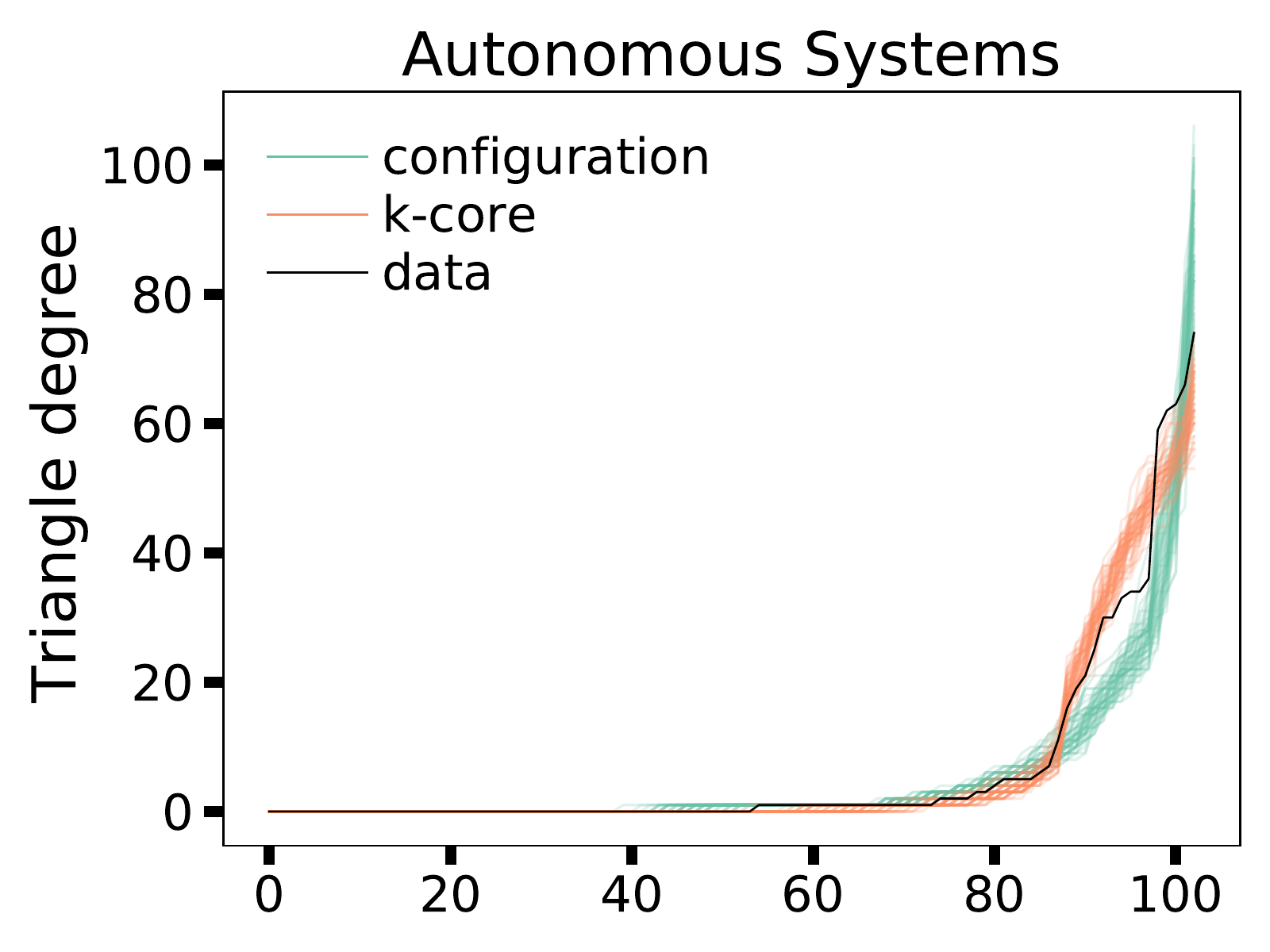}
\includegraphics[width=0.495\columnwidth]{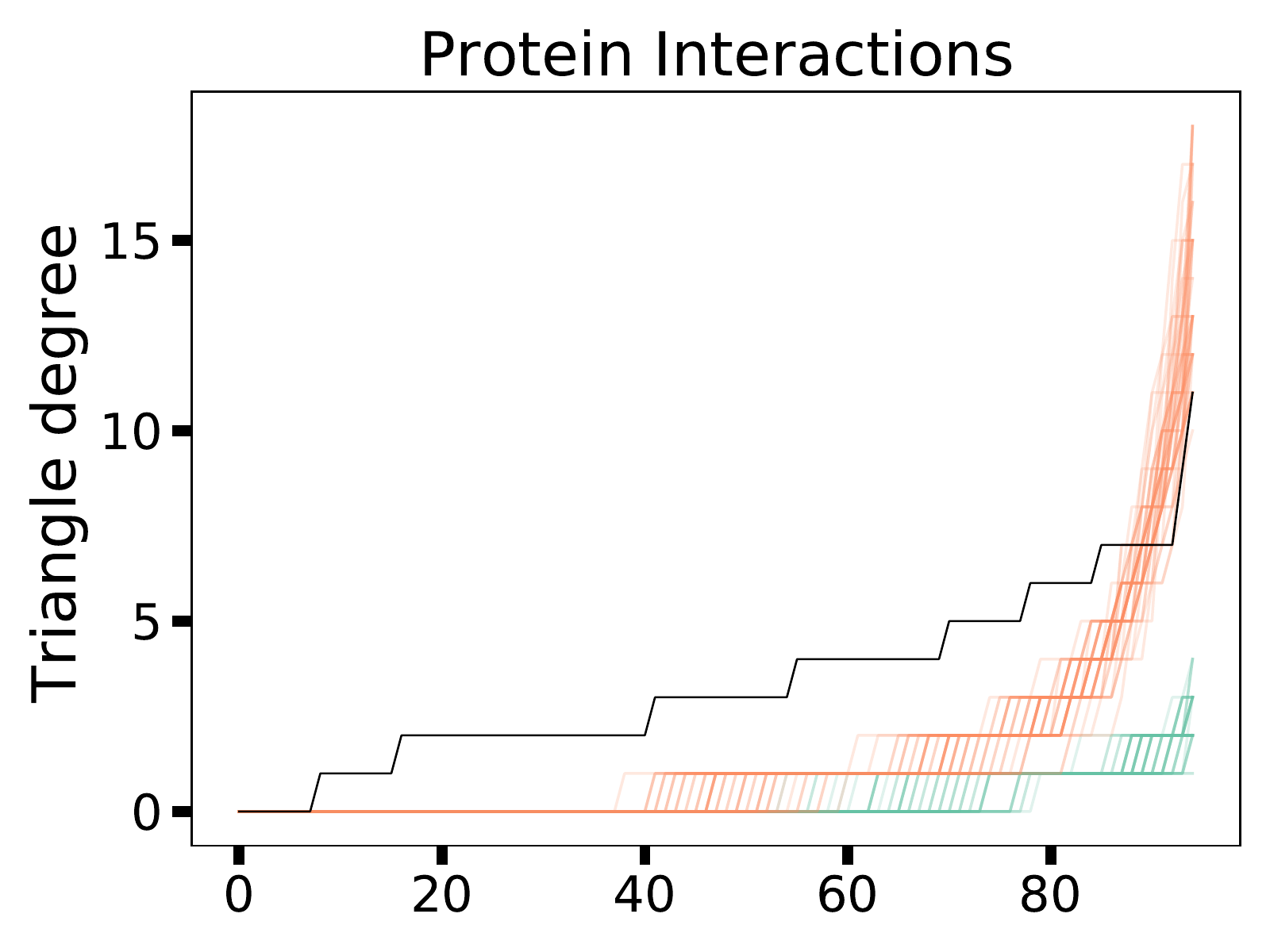}
\includegraphics[width=0.495\columnwidth]{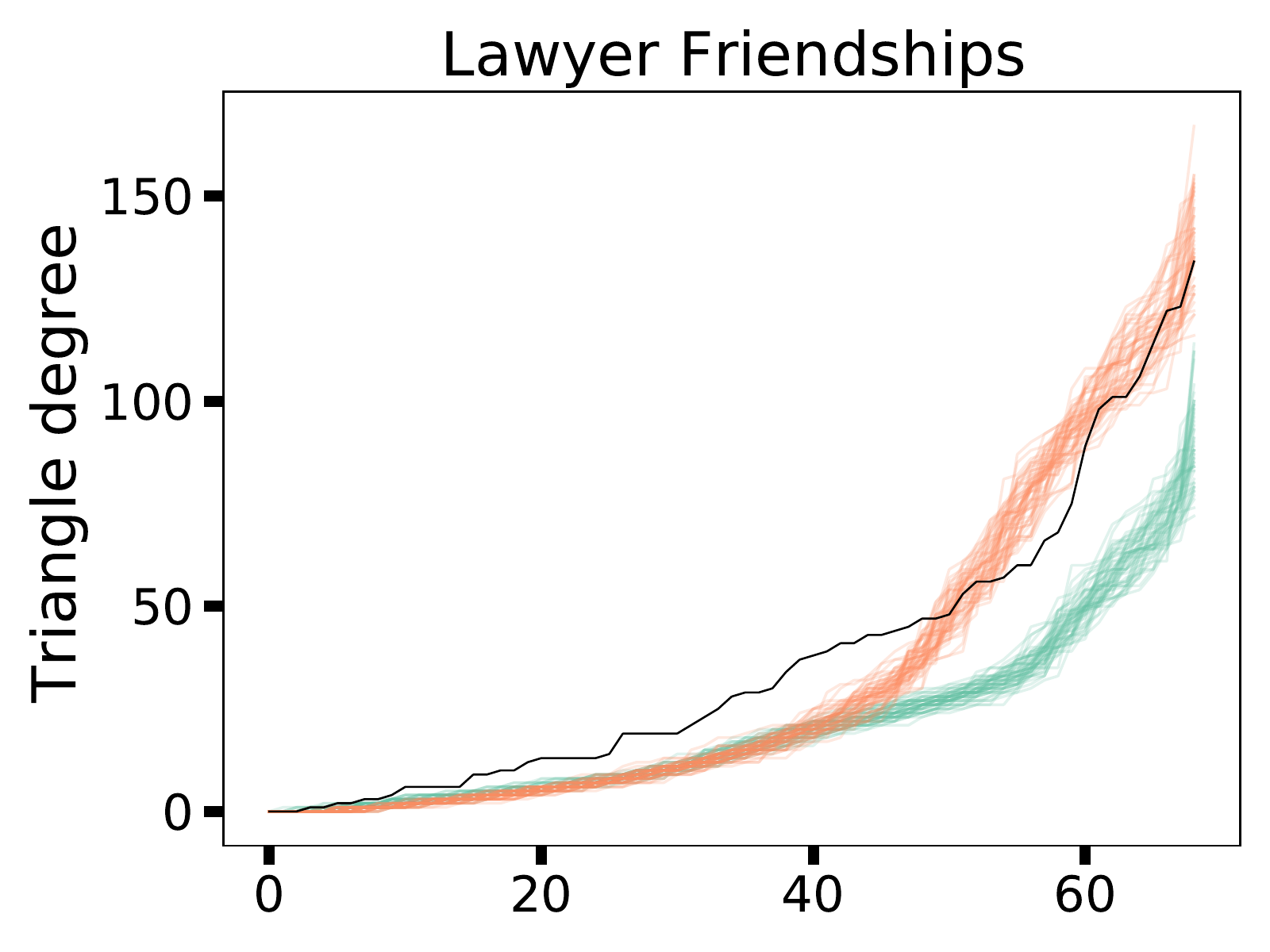}
\includegraphics[width=0.495\columnwidth]{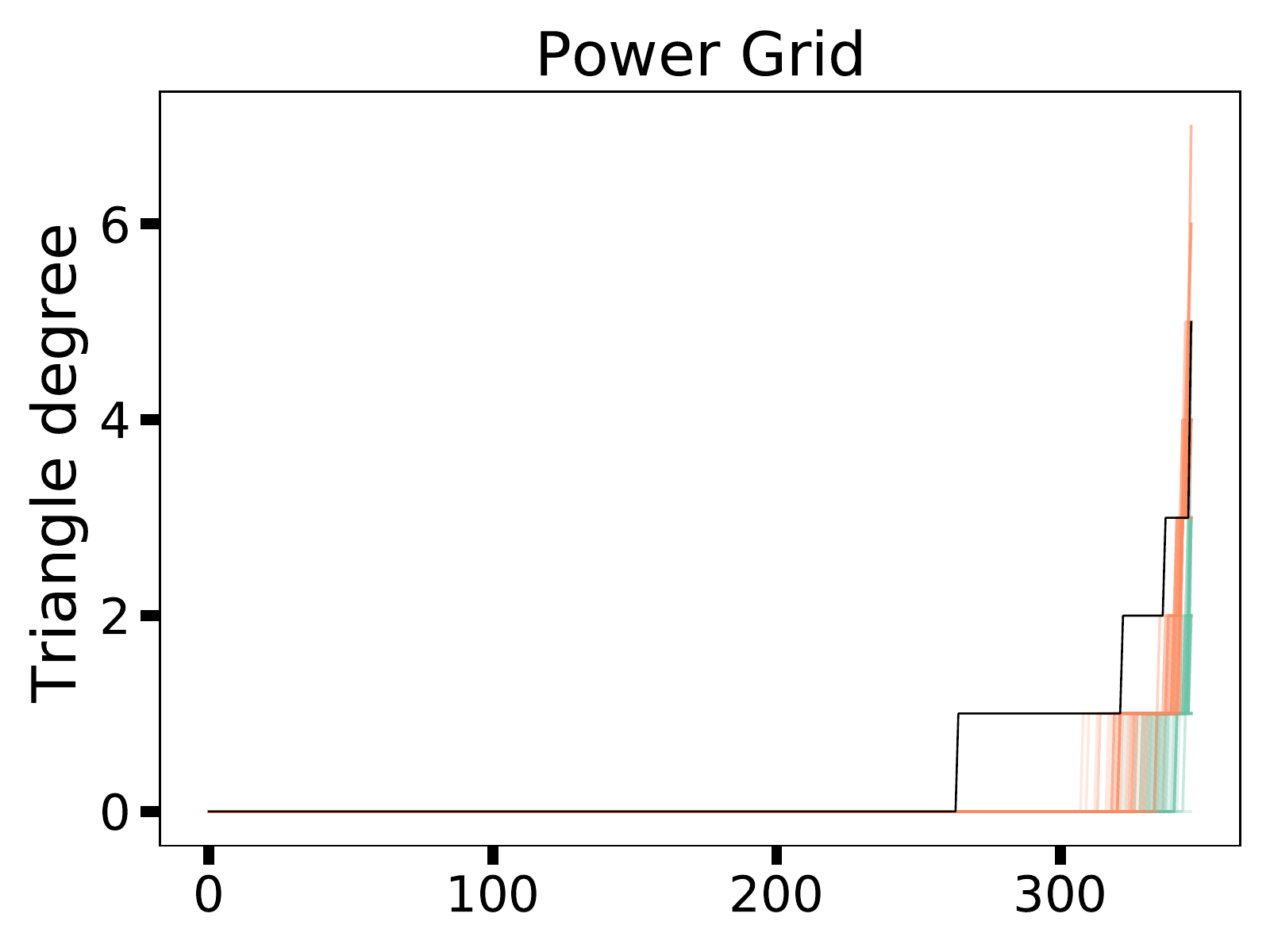}
\Description[Four exponential graphs]{K-core triangle degree is consistently higher than the configuration model and closer to the true data.}
\caption{Triangle degree sequences (given by the number of triangles adjacent to a given node) 
from 50 random samples of graphs with a $k$-core sequence given by a real-world graph dataset
and 50 random samples of graphs with a degree sequence given by a real-world graph dataset.
The $k$-core samples tend to match the triangle sequence more closely.}
\label{fig:tri_seq}
\end{figure}

For our computational experiments here and in a number of the subsequent analyses, we use four graph datasets:
an autonomous systems network~\cite{leskovec2005graphs},
a protein structure network~\cite{milo2004superfamilies},
a friendship network of lawyers working at the same firm~\cite{lazega2001collegial}, and
a power grid~\cite{son_kim_olave-rojas_alvarez-miranda_2018}.
For each dataset, we run our Markov-chain sampler for a number of steps equal to 100 times the number of edges in the graph,
with input $k$-core sequence given by the dataset.
We repeated this 50 times to get 50 random graphs with a prescribed $k$-core distribution.

We then compare the statistics of the resulting graph to the output of the configuration model.
For this, we use 50 samples from a Markov-chain configuration model sampler for vertex-labeled simple graphs,
using the double edge swap procedure described by Fosdick et al.~\cite{fosdick2018configuring}.\footnote{Note that the Markov-chain approach is the standard strategy for generating fixed-degree graphs because we are trying to produce simple graphs; more basic direct approaches yield graphs with self-loops and parallel edges.}

As noted earlier in this section, one weakness of the configuration model is that it destroys local structure, and we observe this even on the small
datasets considered here.
Specifically, the total number of triangles in the configuration model samples is far below the number of triangles
in the corresponding datasets (Figure~\ref{fig:tri_dist}).

The random samples from the prescribed $k$-core sequence have more triangles than those in the configuration model samples.
Moreover, the distribution of the total number of triangles straddles the number of triangles in the autonomous systems dataset.
Thus, the observed number of triangles in this datasets is unsurprising \emph{given the $k$-core sequence}.
In other words, we would not reject the null model of a random graph sampled uniformly at random from the space
of graphs with the given $k$-core sequence, just based on the statistic of the number of triangles.

In addition to the total number of triangles, we also measure the \emph{triangle degree sequence} in these random
samples and compare them to the datasets (Figure~\ref{fig:tri_seq}).
Here, the triangle degree of a node is the number of triangles in which it participates.
We see that the triangle degree sequences given by the $k$-core sequence null model more closely match those of the data.

Taken together, the results of this subsection provide evidence that our $k$-core-based null model
offers a substantially different baseline than the configuration model.
In particular, for the datasets considered here, the core-based null model produces random samples with a larger number of triangles that capture
some of the local structure in the graph.
We will see in the next subsection that this same principle applies for motif analysis more generally.

\subsection*{\small Motif analysis}

A longstanding application of null models for network analysis is the identification
of important or unusual small subgraph patterns called \emph{network motifs}~\cite{milo2002network}.
The main idea is to count the number of occurrences of several small subgraphs in a given dataset as well as in several
random samples from a null model.
``Motifs'' are then subgraphs that appear significantly more or less often than in the null.
Historically, the employed null model is the configuration model~\cite{milo2002network,milo2004superfamilies,fosdick2018configuring}.
Here, we consider both the configuration model and our $k$-core-based model as null models.

\begin{figure}
    \centering
    \includegraphics[width=0.495\columnwidth]{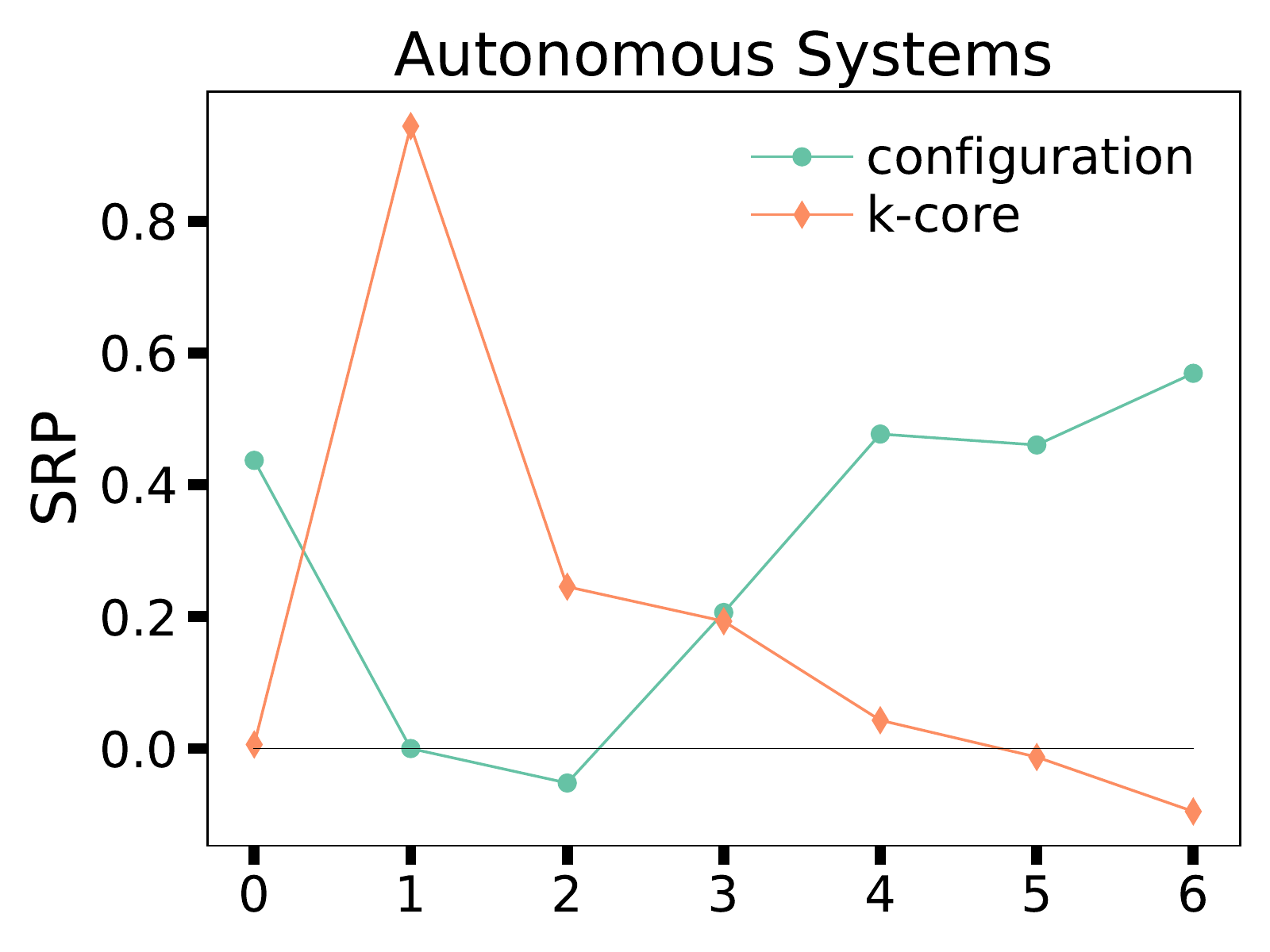}
    \includegraphics[width=0.495\columnwidth]{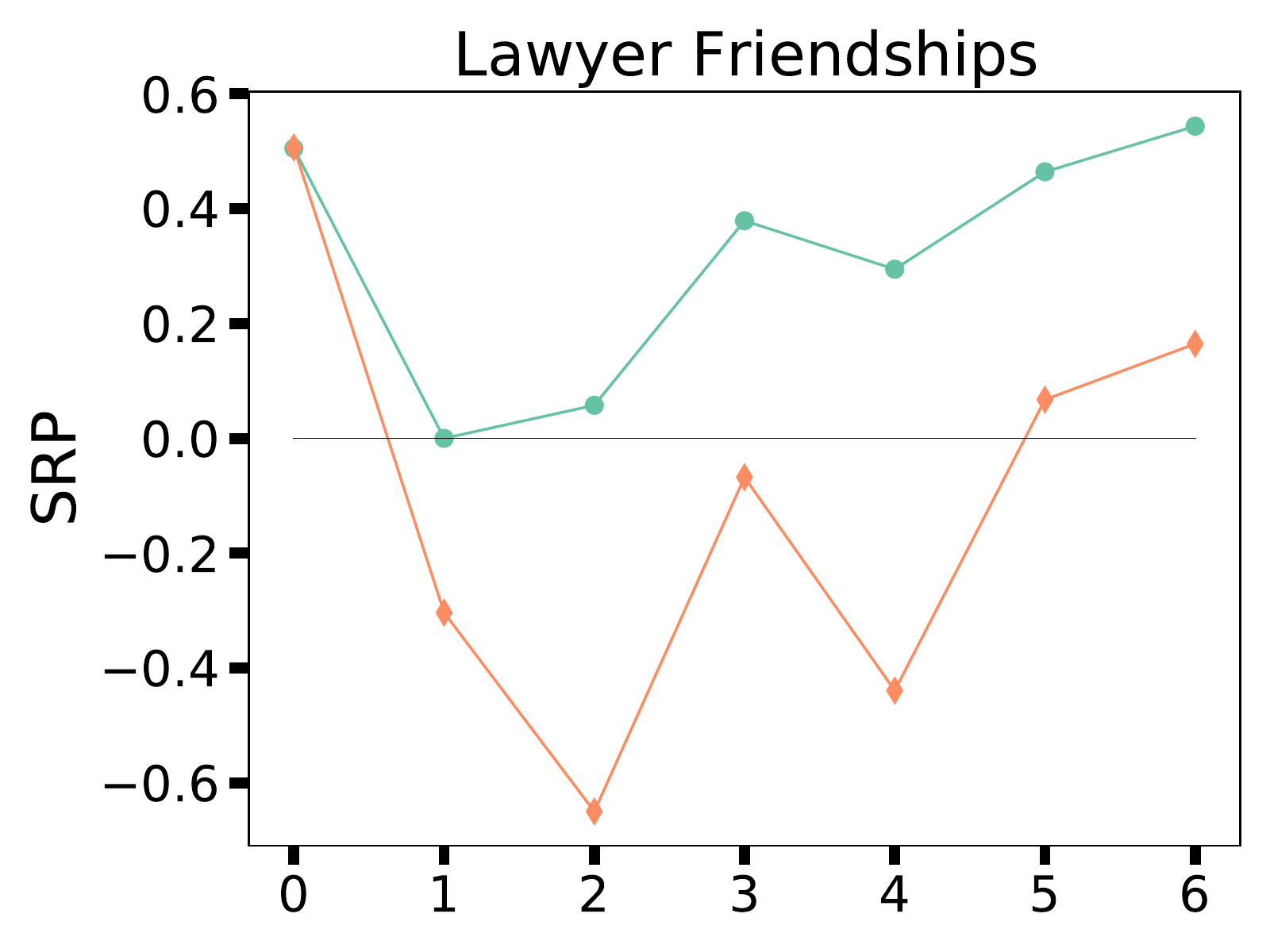}
    \includegraphics[width=0.99\columnwidth]{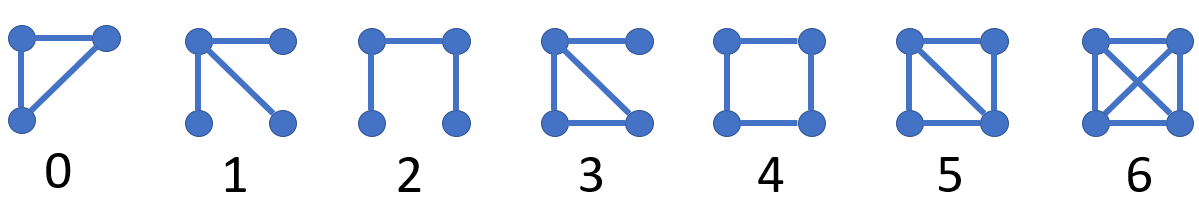}
    \Description[Two line plots]{K-core and configuration model diverge for all 7 subgraphs.}
    \caption{Subgraph ratio profile (SRP) plots under the $k$-core and configuration null models for four node subgraphs and triangles.
      The x-axis in the SRP plots are indexed by the seven subgraphs at the bottom.
    }
    \label{fig:srp}
\end{figure}

In Figure \ref{fig:srp} we consider the results of counting six different motifs consisting of six distinct (non-induced) subgraphs on four nodes each, as well as a motif consisting of the triangle so that we can view the results of the previous subsection in this context as well.
To decide whether the number of copies of a given subgraph appears significantly more or less frequently than in a random baseline, a canonical approach is to the use the {\em subgraph ratio profile (SRP)}, which essentially measures a normalized difference between the frequencies of the subgraph in the real network and in the random baseline. (We refer readers to Milo et al.~\cite{milo2004superfamilies} for the precise definition.)  As a result, a positive SRP for a given subgraph indicates that the subgraph occurs more frequently in the real data than in a random baseline, while a negative SRP indicates that it occurs less frequently.  Positive SRP values are thus taken as evidence that the corresponding subgraph is a meaningfully abundant motif in the network data.

Viewed in this context, we see that the SRP can be defined using any random-graph model that fixes some aspect of the structure of the real network.  While the configuration model that fixes degrees is the standard approach, we can also define SRP values using the core-value model and ask whether we arrive at similar conclusions.
As we see in Figure \ref{fig:srp}, the SRP values based on the core-value model are in fact quite different for two of our datasets, on autonomous systems and the social network on lawyers.\footnote{For power grids and protein networks, there isn't enough meaningful four-node structure to produce clear results using either baseline.}
In particular, we see that many SRP values are on opposite sides of $0$ across the two models, showing that a number of conclusions can change when we move a core-based null model.
Moreover, these changes generally go in the conjectured direction based on the preservation of local structure: if we believe that the core-value model destroys less of the local structure in a network relative to the configuration model, then we would expect lower (and potentially negative) SRP values, and this is what see for many of the subgraphs in Figure~\ref{fig:srp}.
The results thus point to the crucial role in the choice of null model for interpreting these subgraph frequency questions --- a type of issue that becomes feasible to ask given an efficient way to generate null graphs with fixed core-value sequences.

\subsection{Edge-based statistics}

To understand how the core-value model behaves in these types of applications, it is natural to explore some of its basic properties as well.
Perhaps the most fundamental set of properties concern basic counts of edges and degrees.

When sampling based on a $k$-core description given by a dataset, a major difference with the configuration model is that the number of edges in the random sample can differ from those in the dataset.
For a simple example, consider a 4-cycle and the graph obtained by adding one additional edge to the 4-cycle --- all nodes in both graphs have a core value equal to two, but they differ in the number of edges.
Here, we examine the distribution in the number of edges in random samples generated by our algorithm,
where the core-value sequence is generated by a real-world dataset.

\begin{figure}
\includegraphics[width=0.495\columnwidth]{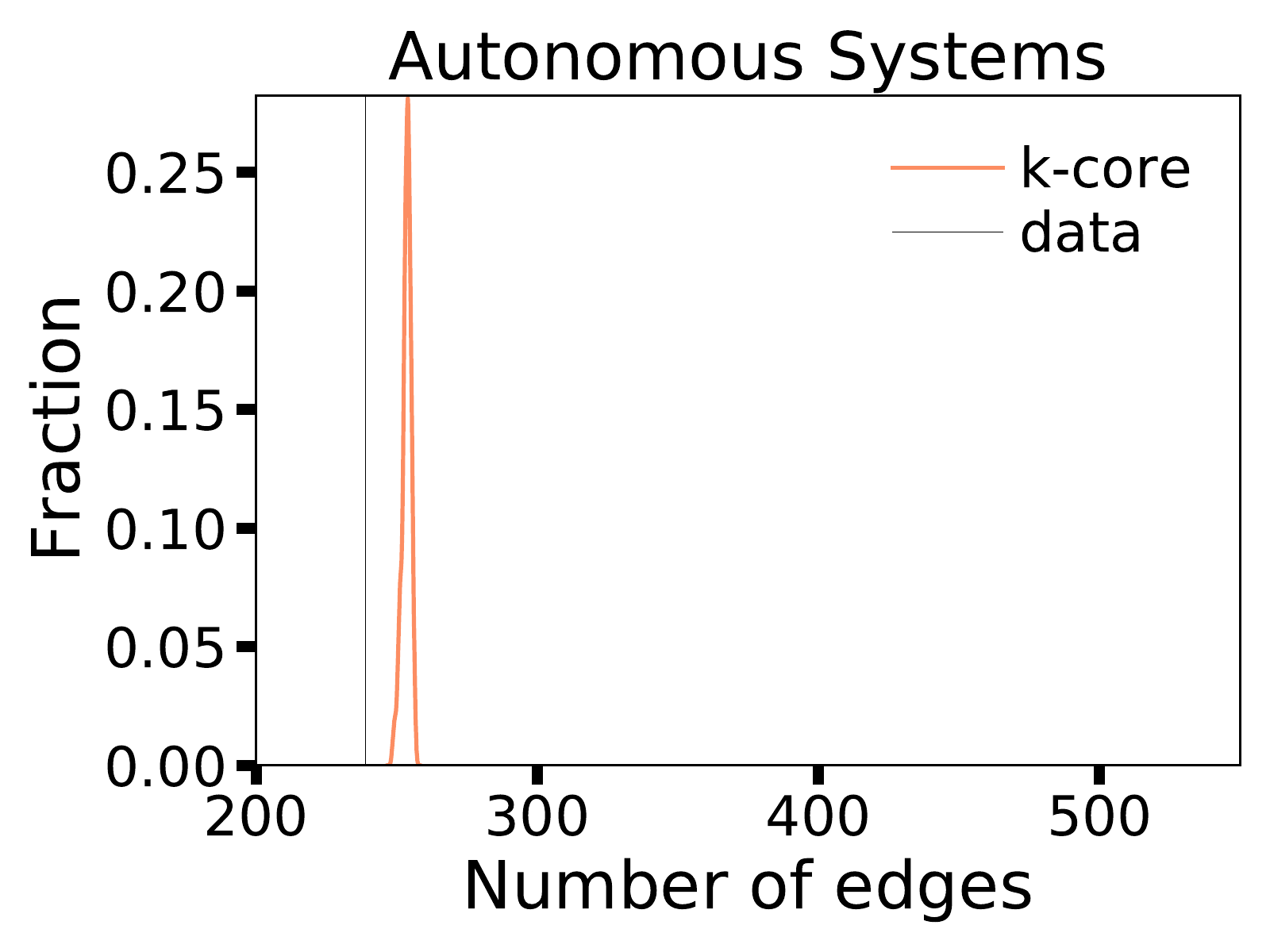}
\includegraphics[width=0.495\columnwidth]{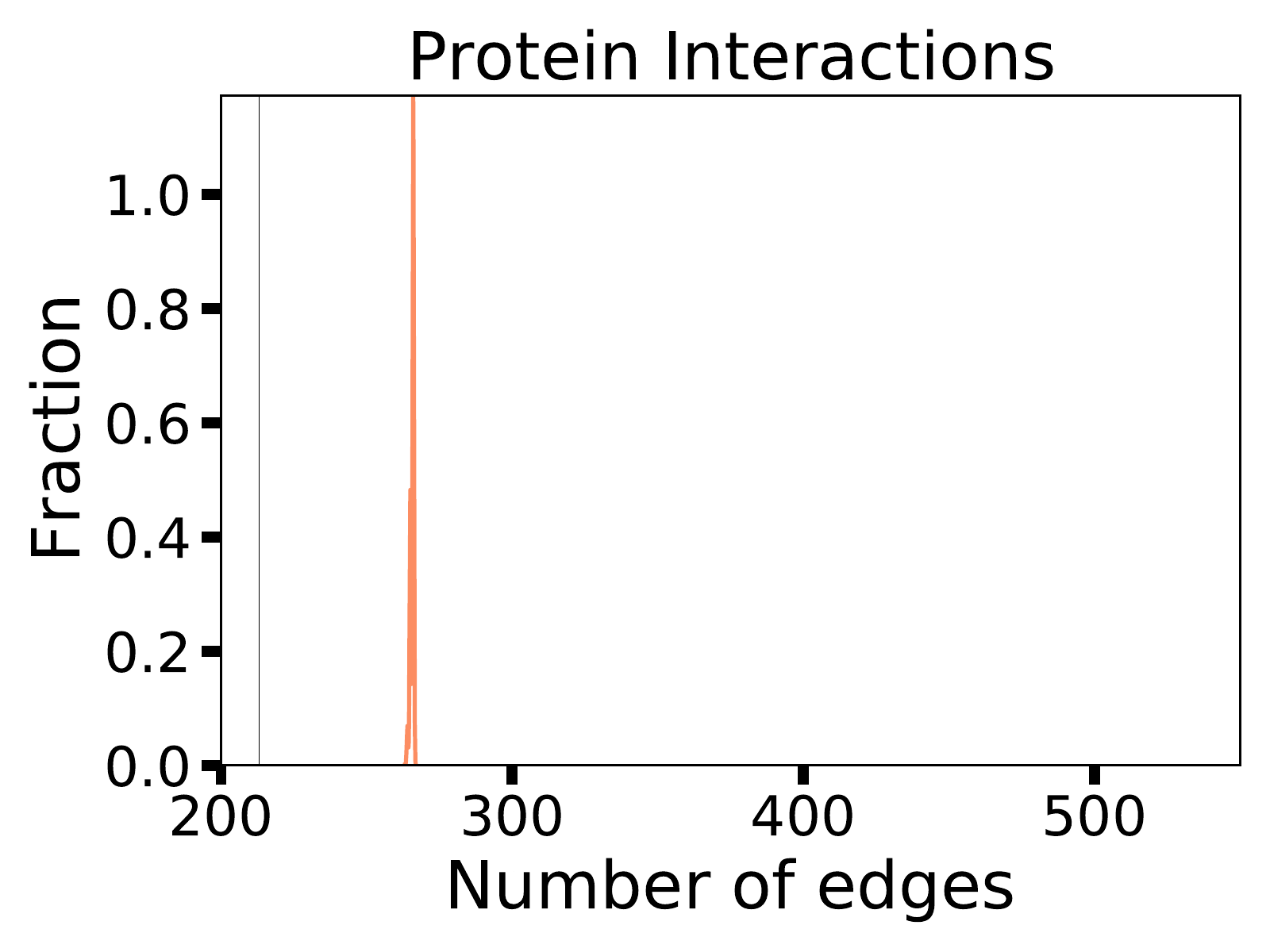}
\includegraphics[width=0.495\columnwidth]{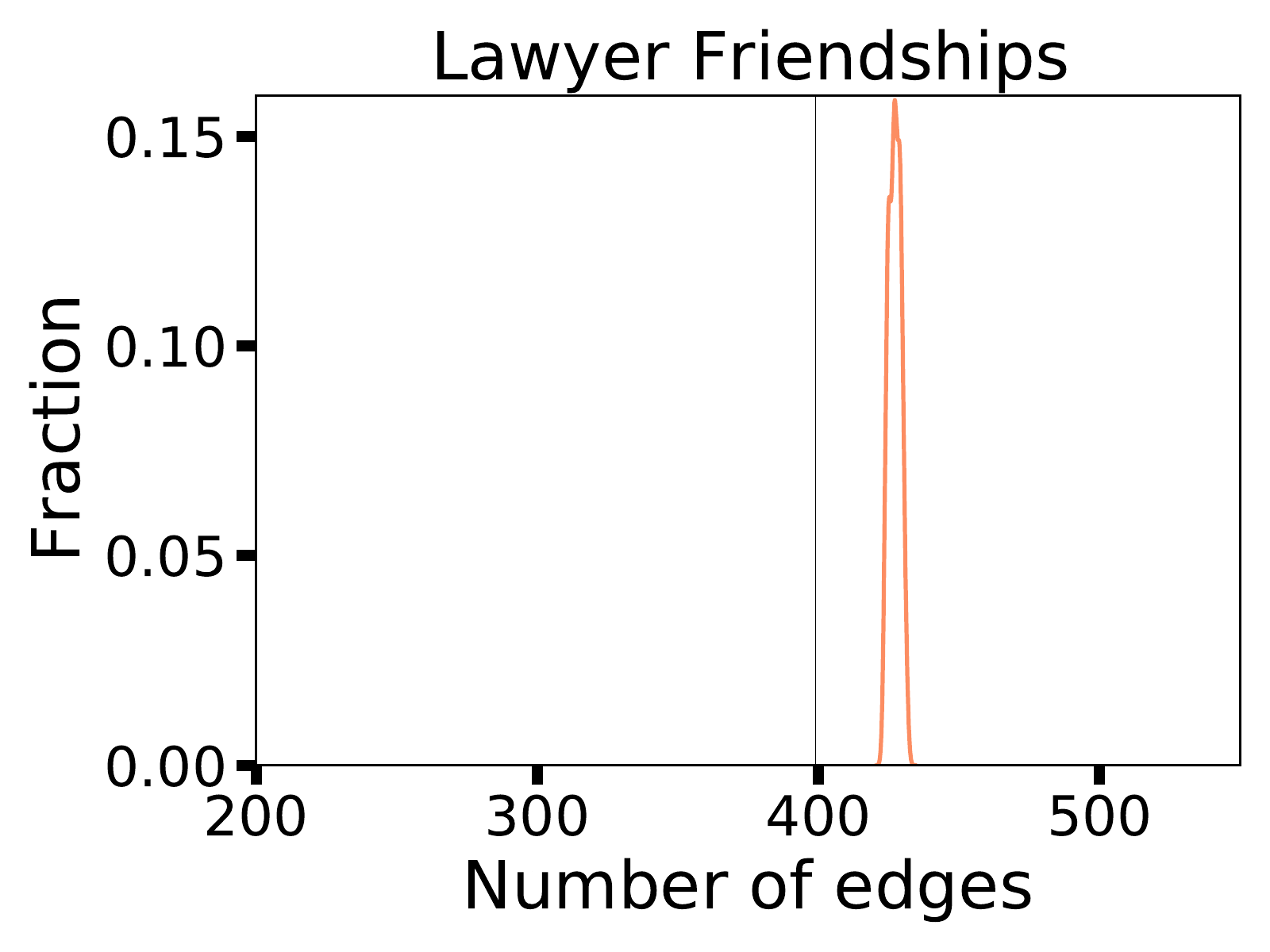}
\includegraphics[width=0.495\columnwidth]{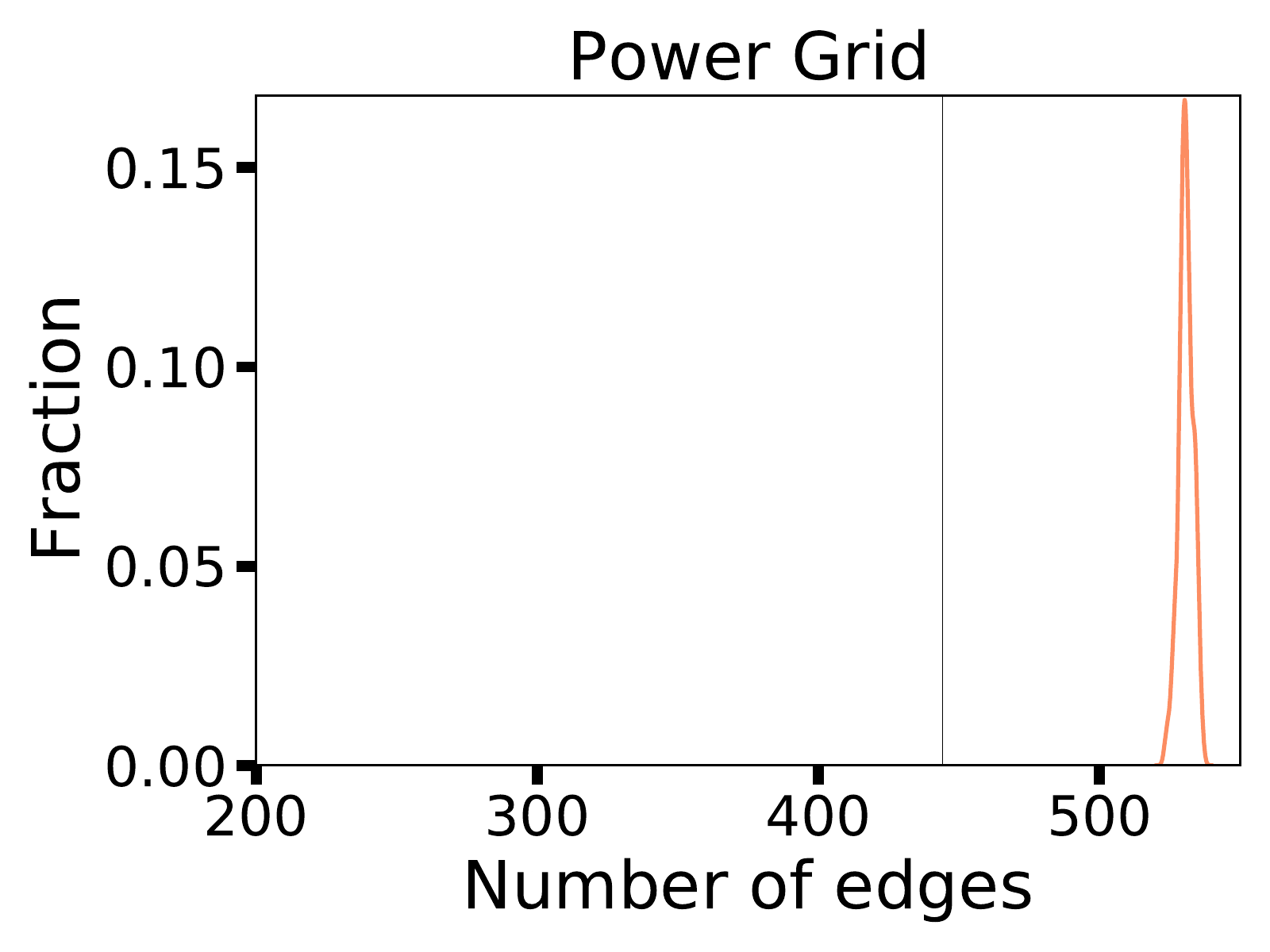}
\Description[Four histograms]{K-core consistently has more edges than the real-world graph.} 
\caption{Distribution of the number of edges from 50 MCMC samples of graphs with a $k$-core sequence given by a real-world graph dataset.
As expected, the number of edges in the random samples is different than in the original data, but the difference is not drastic.}
\label{fig:edge_dist}
\end{figure}

\begin{figure}
\includegraphics[width=0.495\columnwidth]{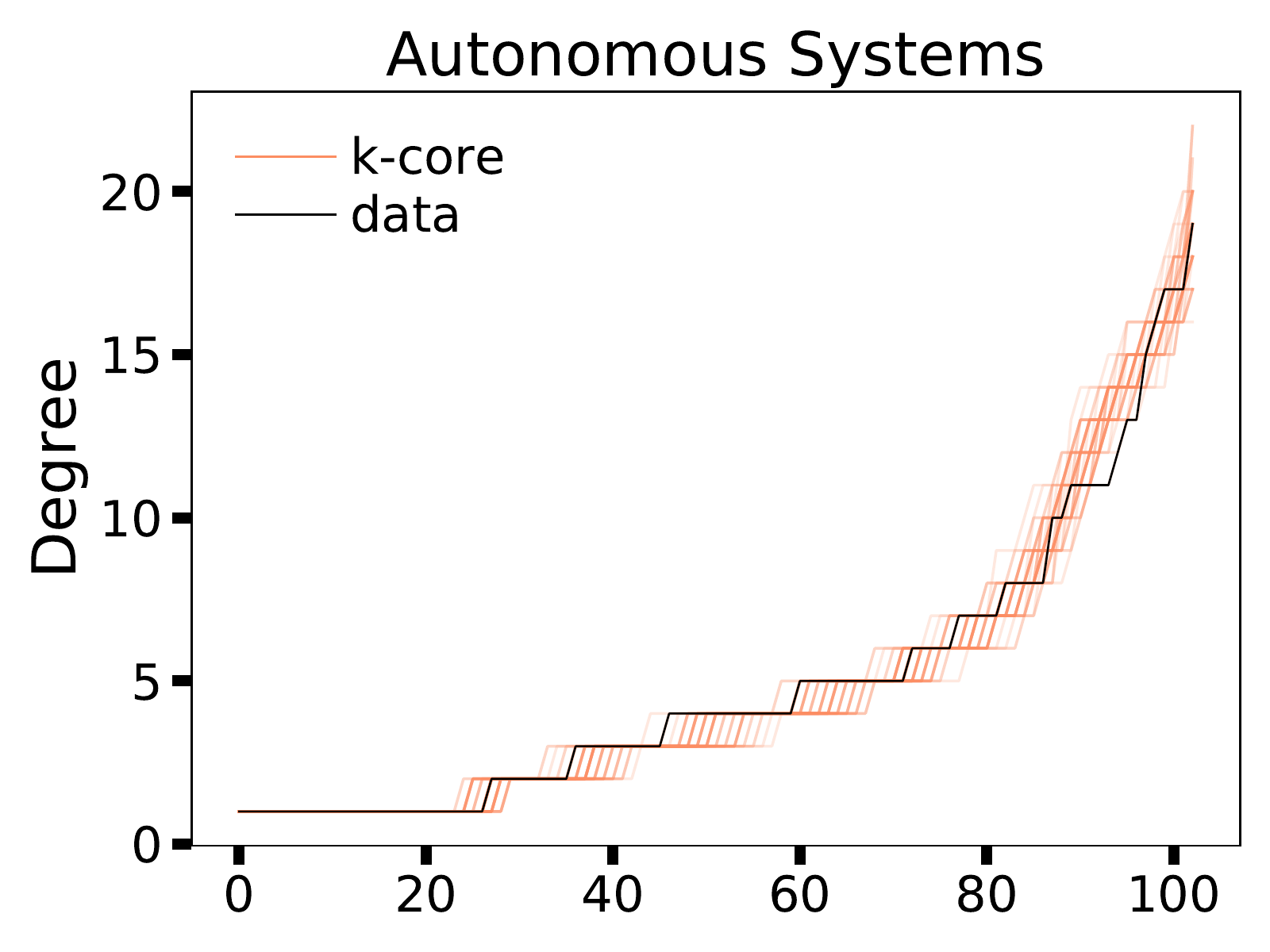}
\includegraphics[width=0.495\columnwidth]{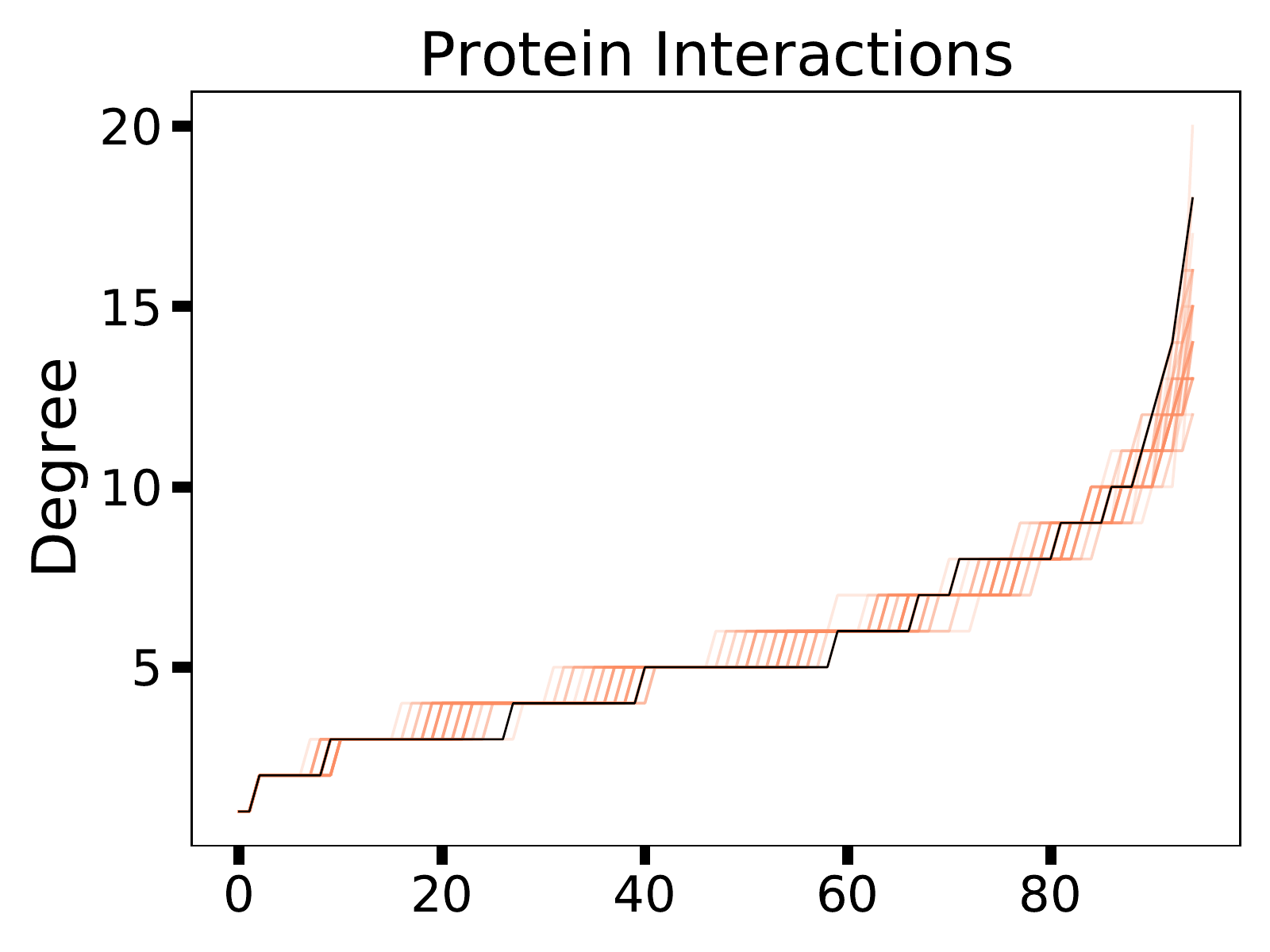}
\includegraphics[width=0.495\columnwidth]{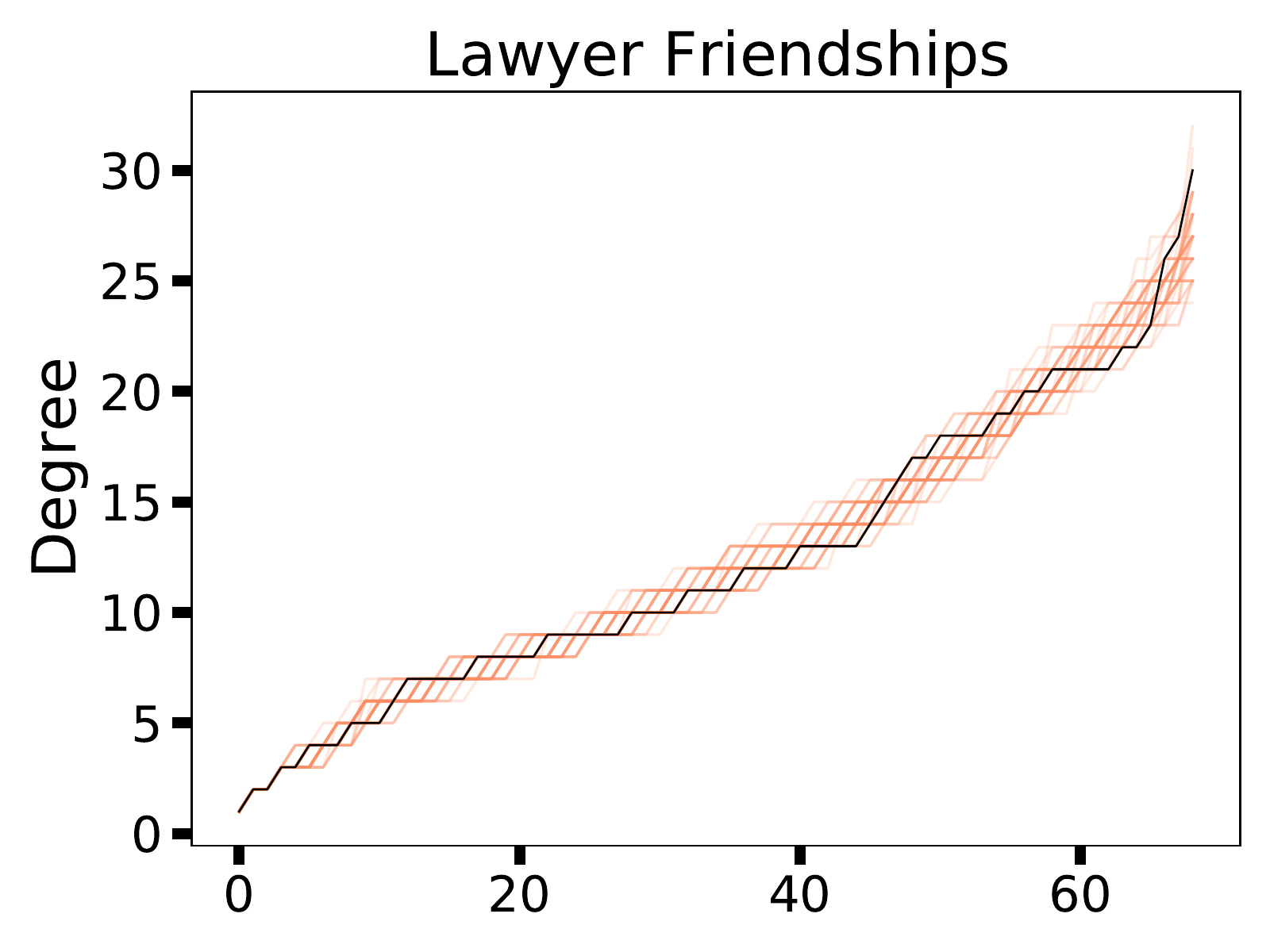}
\includegraphics[width=0.495\columnwidth]{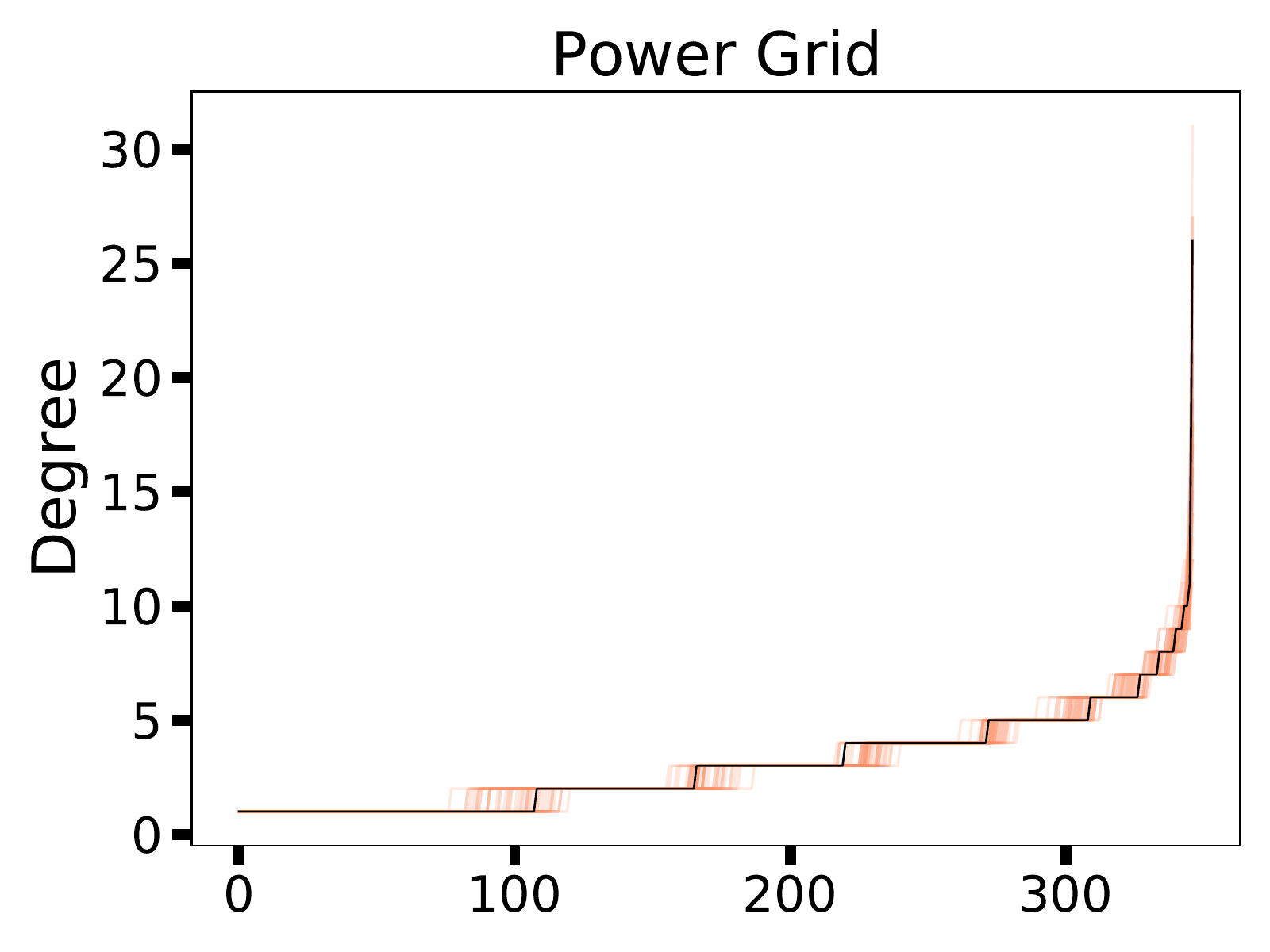}
\Description[Four line plots]{For all four plots, frequency increases with degree.} 
\caption{Degree sequences from 50 MCMC samples of graphs with a $k$-core sequence given by a real-world graph dataset.
The degree sequences of the random samples are similar (but not identical) to the degree sequences in the real-world data.}
\label{fig:deg_seq}
\end{figure}

We use the same datasets and sampling procedure that we employed in the previous subsections.
Figure~\ref{fig:edge_dist} shows the number of edges in the resulting samples.
We see that, for a given dataset, all of the random samples have a number of edges that is greater than or equal to the original data.
Thus, the total number of edges in these datasets over the space of graphs with the same $k$-core sequence is concentrated above the observed number of edges.
At the same time, though, the number of edges in the random sample is not drastically different.

We also compare the degree sequence of the random samples to those in the original data (Figure~\ref{fig:deg_seq}).
The degree sequences largely resemble those in the original data, but are not exactly the same.
Often, the samples from our algorithm produce graphs with a larger maximum degree than the empirical autonomous systems dataset.

\subsection{Attribute-based assortativity}


\begin{table}
  \caption{Network assortativity $r$ with respect to several attributes in the Lawyers dataset. 
  We list the z-score of the assortativity statistic with respect to 50 samples from the configuration
  and $k$-core-based model.
  }
  \label{tab:attrs}
\begin{tabular}{r c c c}
  \toprule
  &     & \multicolumn{2}{c}{z-score} \\
  \cmidrule(lr){3-4} 
  Attribute & $r$ & configuration & $k$-core \\ \midrule
Status             &  0.55  &    21.29 &    3.92       \\
Office Location  &   0.21  &    5.53  &    8.72.   \\
Gender        &      0.12   &   2.50 &       0.31      \\
Law School   &       0.05   &   1.80 &       0.79    \\ 
Type of Practice  &  0.04  &    1.29    &  1.71   \\
\bottomrule
\end{tabular}
\end{table}

As a final investigation, we consider whether or not attribute-based assortativity is preserved 
under the configuration and core-value null models. The lawyers dataset has several
attributes on each node, and we measure the network assortativity $r$~\cite{newman2003mixing}
for status at the firm (partner or associate), office location, gender, law school, and type of practice (litigation or corporate).
Assortativity is positive for all of the attributes, i.e., there is a tendency for edges to appear between two nodes sharing the same attribute
(Table~\ref{tab:attrs}).

As a baseline, we measure the assortativity levels under 50 samples of the configuration model
and the core-value model and compute the same $z$-score as for the motif analysis.
The assortativity scores are higher in the data than in both the null models (all of the $z$-scores in Table~\ref{tab:attrs} are positive).
For example, office location assortativity is overwhelmingly significant under either null.
This is unsurprising, as neither null model is designed to capture mesoscale modular, community, or cluster structure within the network,
and several of the attributes are known to correspond to meaningful cluster structure~\cite{peel2017ground}.

At the same time, evaluating significance based on $z$-scores for some attributes
could lead to different conclusions based on the choice of null model and the desired significance level.
For example, the gender assortativity in the network is $0.12$, which is about 2.5 standard deviations above the mean
with respect to the configuration model, but only 0.31 standard deviations above the mean with respect to the core-value model.
Thus, gender assortativity may seem insignificant under the core-value null but significant under the configuration model null.

\section{Conclusion}

The {\em $k$-core decomposition} is a fundamental graph-theoretic
concept that assigns each node $v$ a {\em core-value} equal to the
largest $c$ such that $v$ belongs to a subgraph of $G$ of minimum degree $c$.
Drawing on this concept,
we have proposed a new family of random graphs that can serve as
a class of null models in network analysis, 
obtained by randomly sampling from the set of
all graphs with a given core-value sequence.
Our sampling method exploits the rich combinatorial structure of
the $k$-core decomposition; we construct a novel Markov chain on
the set of all graphs of a given core-value sequence,
show that the state space is connected with respect to this transition,
and establish that the chain can be used to generate near-uniform
samples from this set of graphs.

The approach opens up a number of intriguing further directions of
potential theoretical and empirical interest.
One question noted earlier is to try establishing bounds on the
mixing rate of the Markov chain we have defined.
Such questions are in general quite challenging, since the mixing
even of simpler chains remains open; we note that many of these chains have
proved valuable for sampling even in the absence of provable guarantees.
A second question, related to our solution of the realizability
question for core-value sequences, is to study extremal questions
over the set of graphs realizing a given core-value sequence;
for example, what is the minimum or maximum number of edges that a graph
with a given core-value sequence can have?
Finally, in a more empirical direction and motivated by our findings
on network motifs, it will be interesting to
characterize the kinds of network properties for which the configuration model
and our core-value model produce systematically different results.
Such comparisons can begin to provide insight into the broader consequences of our choice of null models in network analysis.


\section*{Acknowledgments}
The  authors  thank Haobin Ni for his thoughtful insight.
This research was supported in part by 
ARO Award W911NF19-1-0057, 
ARO MURI, 
NSF Award DMS-1830274, 
a Simons Investigator Award,
a Vannevar Bush Faculty Fellowship,
AFOSR grant FA9550-19-1-0183,
and grants from 
JP Morgan Chase \& Co. and
the MacArthur Foundation.

\bibliographystyle{plain}
\bibliography{main}

\begin{thebibliography}{10}

\bibitem{artzy2005generating}
Yael Artzy-Randrup and Lewi Stone.
\newblock Generating uniformly distributed random networks.
\newblock {\em Physical Review E}, 72(5):056708, 2005.

\bibitem{bender1978asymptotic}
Edward~A Bender and E~Rodney Canfield.
\newblock The asymptotic number of labeled graphs with given degree sequences.
\newblock {\em Journal of Combinatorial Theory, Series A}, 24(3):296--307,
  1978.

\bibitem{bollobas1980probabilistic}
B{\'e}la Bollob{\'a}s.
\newblock A probabilistic proof of an asymptotic formula for the number of
  labelled regular graphs.
\newblock {\em European Journal of Combinatorics}, 1(4):311--316, 1980.

\bibitem{carmi2007model}
Shai Carmi, Shlomo Havlin, Scott Kirkpatrick, Yuval Shavitt, and Eran Shir.
\newblock A model of internet topology using k-shell decomposition.
\newblock {\em Proceedings of the National Academy of Sciences},
  104(27):11150--11154, 2007.

\bibitem{chodrow2020configuration}
Philip~S Chodrow.
\newblock Configuration models of random hypergraphs.
\newblock {\em Journal of Complex Networks}, 8(3):cnaa018, 2020.

\bibitem{choudum1986simple}
Sheshayya~A Choudum.
\newblock A simple proof of the {Erdos-Gallai} theorem on graph sequences.
\newblock {\em Bulletin of the Australian Mathematical Society}, 33(1):67--70,
  1986.

\bibitem{chung2002average}
Fan Chung and Linyuan Lu.
\newblock The average distances in random graphs with given expected degrees.
\newblock {\em Proceedings of the National Academy of Sciences},
  99(25):15879--15882, 2002.

\bibitem{chung2002connected}
Fan Chung and Linyuan Lu.
\newblock Connected components in random graphs with given expected degree
  sequences.
\newblock {\em Annals of combinatorics}, 6(2):125--145, 2002.

\bibitem{chung2004spectra}
Fan Chung, Linyuan Lu, and Van Vu.
\newblock The spectra of random graphs with given expected degrees.
\newblock {\em Internet Mathematics}, 1(3):257--275, 2004.

\bibitem{colomer2013deciphering}
Pol Colomer-de Sim{\'o}n, M~Angeles Serrano, Mariano~G Beir{\'o}, J~Ignacio
  Alvarez-Hamelin, and Mari{\'a}n Bogun{\'a}.
\newblock Deciphering the global organization of clustering in real complex
  networks.
\newblock {\em Scientific reports}, 3:2517, 2013.

\bibitem{dorogovtsev2006k}
Sergey~N Dorogovtsev, Alexander~V Goltsev, and Jose Ferreira~F Mendes.
\newblock K-core organization of complex networks.
\newblock {\em Physical review letters}, 96(4):040601, 2006.

\bibitem{drobyshevskiy2019random}
Mikhail Drobyshevskiy and Denis Turdakov.
\newblock Random graph modeling: A survey of the concepts.
\newblock {\em ACM Computing Surveys (CSUR)}, 52(6):1--36, 2019.

\bibitem{erdos1960graphs}
Paul Erd{\"o}s and Tibor Gallai.
\newblock Graphs with given degrees of vertices, math.
\newblock {\em Mat. Lapok}, 11:264--274, 1960.

\bibitem{fosdick2018configuring}
Bailey~K Fosdick, Daniel~B Larremore, Joel Nishimura, and Johan Ugander.
\newblock Configuring random graph models with fixed degree sequences.
\newblock {\em SIAM Review}, 60(2):315--355, 2018.

\bibitem{gjoka2013}
Minas Gjoka, Maciej Kurant, and Athina Markopoulou.
\newblock 2.5 k-graphs: from sampling to generation.
\newblock In {\em 2013 Proceedings IEEE INFOCOM}, pages 1968--1976. IEEE, 2013.

\bibitem{hakimi1962realizability}
S~Louis Hakimi.
\newblock On realizability of a set of integers as degrees of the vertices of a
  linear graph. i.
\newblock {\em Journal of the Society for Industrial and Applied Mathematics},
  10(3):496--506, 1962.

\bibitem{havel1955remark}
V{\'a}clav Havel.
\newblock A remark on the existence of finite graphs.
\newblock {\em Casopis Pest. Mat.}, 80:477--480, 1955.

\bibitem{kovanen2011temporal}
Lauri Kovanen, M{\'a}rton Karsai, Kimmo Kaski, J{\'a}nos Kert{\'e}sz, and Jari
  Saram{\"a}ki.
\newblock Temporal motifs in time-dependent networks.
\newblock {\em Journal of Statistical Mechanics: Theory and Experiment},
  2011(11):P11005, 2011.

\bibitem{laishram2018measuring}
Ricky Laishram, Ahmet~Erdem Sariy{\"u}ce, Tina Eliassi-Rad, Ali Pinar, and
  Sucheta Soundarajan.
\newblock Measuring and improving the core resilience of networks.
\newblock In {\em Proceedings of the 2018 World Wide Web Conference}, pages
  609--618, 2018.

\bibitem{lazega2001collegial}
Emmanuel Lazega.
\newblock {\em The collegial phenomenon: The social mechanisms of cooperation
  among peers in a corporate law partnership}.
\newblock Oxford University Press on Demand, 2001.

\bibitem{leskovec2005graphs}
Jure Leskovec, Jon Kleinberg, and Christos Faloutsos.
\newblock Graphs over time: densification laws, shrinking diameters and
  possible explanations.
\newblock In {\em Proceedings of the eleventh ACM SIGKDD international
  conference on Knowledge discovery in data mining}, pages 177--187, 2005.

\bibitem{mahadevan2006systematic}
Priya Mahadevan, Dmitri Krioukov, Kevin Fall, and Amin Vahdat.
\newblock Systematic topology analysis and generation using degree
  correlations.
\newblock {\em ACM SIGCOMM Computer Communication Review}, 36(4):135--146,
  2006.

\bibitem{malliaros2020core}
Fragkiskos~D Malliaros, Christos Giatsidis, Apostolos~N Papadopoulos, and
  Michalis Vazirgiannis.
\newblock The core decomposition of networks: Theory, algorithms and
  applications.
\newblock {\em The VLDB Journal}, 29(1):61--92, 2020.

\bibitem{milo2004superfamilies}
Ron Milo, Shalev Itzkovitz, Nadav Kashtan, Reuven Levitt, Shai Shen-Orr, Inbal
  Ayzenshtat, Michal Sheffer, and Uri Alon.
\newblock Superfamilies of evolved and designed networks.
\newblock {\em Science}, 303(5663):1538--1542, 2004.

\bibitem{milo2003uniform}
Ron Milo, Nadav Kashtan, Shalev Itzkovitz, Mark~EJ Newman, and Uri Alon.
\newblock On the uniform generation of random graphs with prescribed degree
  sequences.
\newblock {\em arXiv preprint cond-mat/0312028}, 2003.

\bibitem{milo2002network}
Ron Milo, Shai Shen-Orr, Shalev Itzkovitz, Nadav Kashtan, Dmitri Chklovskii,
  and Uri Alon.
\newblock Network motifs: simple building blocks of complex networks.
\newblock {\em Science}, 298(5594):824--827, 2002.

\bibitem{molloy1995critical}
Michael Molloy and Bruce Reed.
\newblock A critical point for random graphs with a given degree sequence.
\newblock {\em Random structures \& algorithms}, 6(2-3):161--180, 1995.

\bibitem{molloy1998size}
Michael Molloy and Bruce Reed.
\newblock The size of the giant component of a random graph with a given degree
  sequence.
\newblock {\em Combinatorics probability and computing}, 7(3):295--305, 1998.

\bibitem{newman2003mixing}
Mark~EJ Newman.
\newblock Mixing patterns in networks.
\newblock {\em Physical review E}, 67(2):026126, 2003.

\bibitem{newman2001random}
Mark~EJ Newman, Steven~H Strogatz, and Duncan~J Watts.
\newblock Random graphs with arbitrary degree distributions and their
  applications.
\newblock {\em Physical review E}, 64(2):026118, 2001.

\bibitem{orsini2015quantifying}
Chiara Orsini, Marija~M Dankulov, Pol Colomer-de Sim{\'o}n, Almerima Jamakovic,
  Priya Mahadevan, Amin Vahdat, Kevin~E Bassler, Zolt{\'a}n Toroczkai,
  Mari{\'a}n Bogun{\'a}, Guido Caldarelli, et~al.
\newblock Quantifying randomness in real networks.
\newblock {\em Nature communications}, 6(1):1--10, 2015.

\bibitem{peel2017ground}
Leto Peel, Daniel~B Larremore, and Aaron Clauset.
\newblock The ground truth about metadata and community detection in networks.
\newblock {\em Science advances}, 3(5):e1602548, 2017.

\bibitem{sala2010measurement}
Alessandra Sala, Lili Cao, Christo Wilson, Robert Zablit, Haitao Zheng, and
  Zhao~Ben Y.
\newblock Measurement-calibrated graph models for social network experiments.
\newblock In {\em Proceedings of the 19th international conference on World
  wide web}, pages 861--870, 2010.

\bibitem{shen2002network}
Shai~S Shen-Orr, Ron Milo, Shmoolik Mangan, and Uri Alon.
\newblock Network motifs in the transcriptional regulation network of
  escherichia coli.
\newblock {\em Nature genetics}, 31(1):64--68, 2002.

\bibitem{shin2016corescope}
Kijung Shin, Tina Eliassi-Rad, and Christos Faloutsos.
\newblock Corescope: Graph mining using k-core analysis—patterns, anomalies
  and algorithms.
\newblock In {\em 2016 IEEE 16th International Conference on Data Mining
  (ICDM)}, pages 469--478. IEEE, 2016.

\bibitem{son_kim_olave-rojas_alvarez-miranda_2018}
Seung-Woo Son, Heetae Kim, David Olave-Rojas, and Eduardo Álvarez Miranda.
\newblock Node information of chilean power grid with tap, Oct 2018.

\bibitem{sporns2004motifs}
Olaf Sporns and Rolf K{\"o}tter.
\newblock Motifs in brain networks.
\newblock {\em PLoS Biol}, 2(11):e369, 2004.

\bibitem{stanton2012constructing}
Isabelle Stanton and Ali Pinar.
\newblock Constructing and sampling graphs with a prescribed joint degree
  distribution.
\newblock {\em Journal of Experimental Algorithmics (JEA)}, 17:3--1, 2012.

\bibitem{wilson1996generating}
David~Bruce Wilson.
\newblock Generating random spanning trees more quickly than the cover time.
\newblock In Gary~L. Miller, editor, {\em Proceedings of the Twenty-Eighth
  Annual {ACM} Symposium on the Theory of Computing, Philadelphia,
  Pennsylvania, USA, May 22-24, 1996}, pages 296--303. {ACM}, 1996.

\bibitem{young2017construction}
Jean-Gabriel Young, Giovanni Petri, Francesco Vaccarino, and Alice Patania.
\newblock Construction of and efficient sampling from the simplicial
  configuration model.
\newblock {\em Physical Review E}, 96(3):032312, 2017.

\end{thebibliography}

\end{document}